\documentclass{emulateapj}
\usepackage{graphicx}
\usepackage{amssymb,amsfonts,amsmath,amstext,amsgen,amsopn,amsxtra,indentfirst,times}

\begin{document}

\title{Analytical Models of Exoplanetary Atmospheres. \\II. Radiative Transfer via the Two-Stream Approximation}
\author{Kevin Heng\altaffilmark{1}}
\author{Jo\~{a}o M. Mendon\c{c}a\altaffilmark{1}}
\author{Jae-Min Lee\altaffilmark{1,2}}
\altaffiltext{1}{University of Bern, Center for Space and Habitability, Sidlerstrasse 5, CH-3012, Bern, Switzerland.  Email: kevin.heng@csh.unibe.ch, joao.mendonca@csh.unibe.ch}
\altaffiltext{2}{University of Z\"{u}rich, Institute for Computational Science, Winterthurerstrasse 190, CH-8057, Z\"{u}rich, Switzerland.  Email: lee@physik.uzh.ch}

\begin{abstract}
We present a comprehensive analytical study of radiative transfer using the method of moments and include the effects of non-isotropic scattering in the coherent limit.  Within this unified formalism, we derive the governing equations and solutions describing two-stream radiative transfer (which approximates the passage of radiation as a pair of outgoing and incoming fluxes), flux-limited diffusion (which describes radiative transfer in the deep interior) and solutions for the temperature-pressure profiles.  Generally, the problem is mathematically under-determined unless a set of closures (Eddington coefficients) is specified.  We demonstrate that the hemispheric (or hemi-isotropic) closure naturally derives from the radiative transfer equation if energy conservation is obeyed, while the Eddington closure produces spurious enhancements of both reflected light and thermal emission.  We concoct recipes for implementing two-stream radiative transfer in stand-alone numerical calculations and general circulation models.  We use our two-stream solutions to construct toy models of the runaway greenhouse effect.  We present a new solution for temperature-pressure profiles with a non-constant optical opacity and elucidate the effects of non-isotropic scattering in the optical and infrared.  We derive generalized expressions for the spherical and Bond albedos and the photon deposition depth.  We demonstrate that the value of the optical depth corresponding to the photosphere is not always 2/3 (Milne's solution) and depends on a combination of stellar irradiation, internal heat and the properties of scattering both in optical and infrared.  Finally, we derive generalized expressions for the total, net, outgoing and incoming fluxes in the convective regime.
\end{abstract}


\section{Introduction}

\begin{table}
\label{tab:symbols}
\begin{center}
\caption{Commonly Used Symbols}
\begin{tabular}{lcc}
\hline
\hline
Name & Units & Meaning \\
\hline
$\mu$ & --- & cosine of zenith angle \\
$\bar{\mu}$ & --- & characteristic or mean value of $\mu$ \\
${\cal T}$ & --- & transmission function or transmissivity$^\dagger$ \\
$\omega_0$ & --- & single-scattering albedo$^\dagger$ \\
$g_0$ & --- & scattering asymmetry factor$^\dagger$ \\
$\beta_0$ & --- & $\equiv \sqrt{\frac{1-\omega_0}{1-\omega_0 g_0}}$; scattering parameter$^\dagger$ \\
$\zeta_\pm$ & --- & $\equiv \left( 1 \pm \beta_0 \right)/2$; coupling coefficients$^\dagger$ \\
$\beta_{\rm S_0}$ & --- & shortwave/optical scattering parameter \\
$\beta_{\rm L_0}$ & --- & longwave/infrared scattering parameter \\
$A_g$ & --- & geometric albedo$^\dagger$ \\
$A_{\rm s}$ & --- & spherical albedo$^\dagger$ \\
$A_{\rm B}$ & --- & Bond albedo \\
$g$ & cm s$^{-2}$ & surface gravity of exoplanet \\
$a$ & cm & spatial separation (exoplanet and star) \\
\hline
$\tau_0$ & --- & optical depth$^\dagger$ \\
$\tau$ & --- & slant optical depth$^\dagger$ \\
$\kappa$ & cm$^2$ g$^{-1}$ & total/extinction opacity $^\dagger$ \\
$\kappa_{\rm a}$ & cm$^2$ g$^{-1}$ & absorption opacity $^\dagger$ \\
$m$ & g cm$^{-2}$ & column mass \\
$I$ & erg cm$^{-3}$ s$^{-1}$ sr$^{-1}$ & intensity$^\dagger$ \\
$J$ & erg cm$^{-3}$ s$^{-1}$ & total intensity$^\dagger$ \\
$F_\uparrow$ & erg cm$^{-3}$ s$^{-1}$ & outgoing flux$^\dagger$ \\
$F_\downarrow$ & erg cm$^{-3}$ s$^{-1}$ & incoming flux$^\dagger$ \\
$F_+$ & erg cm$^{-3}$ s$^{-1}$ & total flux$^\dagger$ \\
$F_-$ & erg cm$^{-3}$ s$^{-1}$ & net flux$^\dagger$ \\
$B$ & erg cm$^{-3}$ s$^{-1}$ sr$^{-1}$ & Planck function$^\dagger$ \\
\hline
$\kappa_{\rm R}$ & cm$^2$ g$^{-1}$ & Rosseland-mean opacity \\
$\kappa_{\rm S}$ & cm$^2$ g$^{-1}$ & shortwave/optical opacity \\
$n$ & --- & shortwave opacity index \\
$\kappa_{\rm L}$ & cm$^2$ g$^{-1}$ & longwave/infrared opacity \\
$\omega_{\rm S_0}$ & --- & single-scattering albedo (shortwave) \\
$\omega_{\rm L_0}$ & --- & single-scattering albedo (longwave) \\
$g_{\rm S_0}$ & --- & asymmetry factor (shortwave) \\
$g_{\rm L_0}$ & --- & asymmetry factor (longwave) \\
${\cal J}$ & erg cm$^{-2}$ s$^{-1}$ & total intensity (all wavelengths) \\
${\cal F}_\uparrow$ & erg cm$^{-2}$ s$^{-1}$ & outgoing flux (all wavelengths) \\
${\cal F}_\downarrow$ & erg cm$^{-2}$ s$^{-1}$ & incoming flux (all wavelengths) \\
${\cal F}_+$ & erg cm$^{-2}$ s$^{-1}$ & total flux (all wavelengths) \\
${\cal F}_-$ & erg cm$^{-2}$ s$^{-1}$ & net flux (all wavelengths) \\
${\cal E}_i$ & --- & $i$-th order exponential integral \\
\hline
$T$ & K & temperature \\
$\bar{T}$ & K & global-mean temperature \\
$T_{\rm irr}$ & K & irradiation temperature \\
$T_\star$ & K & effective stellar temperature \\
$T_{\rm int}$ & K & internal temperature \\
\hline
\hline
\end{tabular}\\
$\dagger$: quantity is wavelength-dependent.\\
\end{center}
\end{table}

The ability of astronomers to measure the spectral energy distributions and transmission spectra of exoplanetary atmospheres has inspired theoretical efforts to model, interpret and predict their spectral and thermal structures.  Techniques range from studying atmospheres in radiative and/or chemical equilibrium (e.g., \citealt{burrows08,fortney10}) to the inference of chemical composition and thermal structure based solely on the data (e.g., \citealt{benneke12,lee12,line13}).  Global climate models have been adapted to study the radiation hydrodynamics of exoplanetary atmospheres (e.g., \citealt{showman09,hmp11,rm12}).  This surge of interest motivates a careful re-examination of the assumptions and techniques used in radiative transfer, since the exoplanetary atmospheres accessible to astronomical measurement reside in non-Solar-System-centric regimes.\footnote{Presently, these are highly-irradiated exoplanets with temperatures $\sim 800$--3000 K.}

Central to these theoretical efforts is a simple, general and fast technique to compute radiative transfer known as the ``two-stream approximation" \citep{chandra60,mw80,gy89,toon89,mm99}.  It solves the moments of the radiative transfer equation and treats the passage of radiation through an atmosphere as a pair of outgoing and incoming rays.  It is versatile enough to be used in stand-alone calculations of atmospheres in radiative equilibrium, retrieval calculations or coupled to three-dimensional general circulation models.  Related to this technique are analytical calculations of temperature-pressure profiles \citep{hubeny03,hansen08,guillot10,hhps12,robinson12,pg14}, which allow one to develop intuition for the thermal structure of an atmosphere.  While the two-stream treatment itself is not novel, it comes in several flavors \citep{pierrehumbert}, is often tuned toward studying the Earth, Solar System, brown dwarfs or stars and there is a need to elucidate the assumptions involved so that we can harness it to study exoplanetary atmospheres.

The over-arching goal of the present study is to construct a unified formalism for calculating radiative transfer and analytical temperature-pressure profiles.  We examine the two-stream radiative transfer method applied to atmospheres, flux-limited diffusion as a description of radiative transfer in the deep interior of exoplanets and temperature-pressure profiles, all in the limit of non-isotropic, coherent scattering.  Each of these techniques has previously been studied separately, but not in a unified manner using a self-consistent set of governing equations.  Since we are dealing with moments of the radiative transfer equation, a set of closures (Eddington coefficients) is needed such that the problem is not mathematically under-determined.  One of our goals is to derive a self-consistent set of closures.  By distinguishing between the total, net, outgoing and incoming fluxes, we resolve several inconsistencies lingering in the literature.  

In \S\ref{sect:isotropic}, we revisit two-stream radiative transfer in the limit of isotropic, coherent scattering.  In the process, we demonstrate that the hemispheric closure naturally derives from the radiative transfer equation, while recommending that the Eddington closure not be used.  In \S\ref{sect:non_isotropic}, we examine non-isotropic, coherent scattering and describe a transition to flux-limited diffusion in the deep interior.  In \S\ref{sect:tp}, we use our findings in \S\ref{sect:isotropic} and \S\ref{sect:non_isotropic} to derive analytical temperature-pressure profiles with non-isotropic, coherent scattering and a non-constant optical or shortwave opacity.  In \S\ref{sect:apps}, we apply our unified formalism to studying other closures, concocting recipes for computing two-stream radiative transfer, calculating albedo spectra and temperature-pressure profiles, generalizing Milne's solution and constructing toy models of the runaway greenhouse effect.  In \S\ref{sect:discussion}, we compare the current study to previous ones and discuss the implications of our findings.  In Appendix {\ref{append:convective}, we derive generalized analytical expressions for the total, net, outgoing and incoming fluxes in the convective regime.  In Appendix \ref{append:direct}, we demonstrate that direct analytical solutions of the radiative transfer equation are only obtainable in the limit of pure absorption.  Table 1 lists the commonly used symbols in this study, while Tables 2 and 3 summarize the closures used and a comparison of the different closures in the literature, respectively.

The present paper is the second in a series of analytical studies that aim to re-examine and generalize the theoretical formalism used in planetary atmospheres.  The first paper studied atmospheric dynamics via the shallow water approximation \citep{hw14}.

\section{Two-Stream Radiative Transfer: Isotropic, Coherent Scattering}
\label{sect:isotropic}

\subsection{Radiative Transfer Equation}

The radiative transfer equation for a plane-parallel, static atmosphere may be stated in a compact form \citep{chandra60,mihalas,gy89,mm99},
\begin{equation}
\mu \frac{\partial I}{\partial \tau_0} = I - S,
\label{eq:rt}
\end{equation}
where $\mu \equiv \cos\theta$ is the cosine of the zenith angle, $I$ is the wavelength-dependent intensity, $\tau_0$ is the wavelength-dependent optical depth and $S$ is the source function.  The source function hides the complexity associated with scattering and thermal emission.  The zenith angle $\theta$ is the angle between an incoming or outgoing ray and the normal to the plane.  Note that we have defined $\tau_0=0$ at the top of the atmosphere.

Generally, it is challenging to obtain analytical solutions of the radiative transfer by directly solving for $I$.  This is possible only in the limit of pure absorption (see Appendix \ref{append:direct}).  Instead, one solves moments of equation (\ref{eq:rt}), which requires us to define the moments of $I$.  The two-stream equations are essentially the first and second moments of equation (\ref{eq:rt}) \citep{mw80}.

\subsection{Moments of the Intensity}

The zeroth, first and second moments of the intensity are
\begin{equation}
\begin{split}
J_\uparrow &\equiv \int^{2\pi}_0 \int^1_0 I ~d\mu ~d\phi, \\
J_\downarrow &\equiv \int^{2\pi}_0 \int^0_{-1} I ~d\mu ~d\phi, \\
F_\uparrow &\equiv \int^{2\pi}_0 \int^1_0 \mu I ~d\mu ~d\phi, \\
F_\downarrow &\equiv \int^{2\pi}_0 \int^0_{-1} \mu I ~d\mu ~d\phi, \\
K_\uparrow &\equiv \int^{2\pi}_0 \int^1_0 \mu^2 I ~d\mu d\phi, \\
K_\downarrow &\equiv \int^{2\pi}_0 \int^0_{-1} \mu^2 I ~d\mu d\phi.
\end{split}
\label{eq:moments}
\end{equation}
The outgoing flux and incoming flux are given by $F_\uparrow$ and $F_\downarrow$, respectively.  Note that the total intensity ($J$), total flux ($F_+$) and net flux ($F_-$), as well as the second moments ($K_\pm$), are given by
\begin{equation}
\begin{split}
J &\equiv J_\uparrow + J_\downarrow, \\
F_\pm &\equiv F_\uparrow \pm F_\downarrow,\\
K_\pm &\equiv K_\uparrow \pm K_\downarrow,
\end{split}
\label{eq:moments2}
\end{equation}
while $E = J/c$ is the energy density, where $c$ is the speed of light.  In a departure from the traditional approach, we have defined total quantities (integrated over one or both hemispheres) and not mean ones (which are further divided by $4\pi$).

\begin{table}
\label{tab:closures}
\begin{center}
\caption{Closures Adopted (Eddington Coefficients)}
\begin{tabular}{lc}
\hline
\hline
Symbol & Meaning \\
\hline
$\epsilon_+ \equiv F_\uparrow/J_\uparrow$ &  first Eddington coefficient (outgoing) \\
$\epsilon_- \equiv F_\downarrow/J_\downarrow$ & first Eddington coefficient (incoming) \\
$\epsilon \equiv F_+/J$ & first Eddington coefficient \\
$\epsilon_2 \equiv K_-/F_+$ & second Eddington coefficient \\
$\epsilon_3 \equiv K_-/J$ & third Eddington coefficient \\
$\epsilon_{\rm S} \equiv K_{\rm S}/J_{\rm S}$ & third Eddington coefficient (shortwave) \\
$\epsilon_{\rm L} \equiv F_{\rm L}/J_{\rm L}$ & first Eddington coefficient (longwave) \\
$\epsilon_{\rm L_3} \equiv K_{\rm L}/J_{\rm L}$ & third Eddington coefficient (longwave) \\
\hline
\hline
Closure & Assumption or Constraint \\
\hline
$\epsilon_\pm = \epsilon_+ = \epsilon_-$ & symmetry between hemispheres \\
$\bar{\mu} = \bar{\mu}_+ = -\bar{\mu}_-$ & symmetry between hemispheres \\
$\epsilon = \epsilon_\pm$ & conservation of energy \\
$\epsilon_\pm = 1/2$ & correct blackbody emission for opaque atmosphere \\
$\epsilon_2 = F_+/2F_-$ & reproduces isotropic limit \\
$\epsilon_3 = 1/3$ & deep atmosphere limit \\
$\epsilon_{\rm S} = \mu^2$ & reproduces Beer's law \\
$\epsilon_{\rm L} = 3/8$ & equal to $\epsilon^2/2\epsilon_3$ (consistency with other closures) \\
$\epsilon_{\rm L_3} =1/3$ & correspondence to $\epsilon_3$ \\
\hline
\hline
\end{tabular}\\
\end{center}
\end{table}

\subsection{Deriving the Two-Stream Form}

The radiative transfer equation with isotropic, coherent scattering is described by \citep{mihalas}
\begin{equation}
\mu \frac{\partial I}{\partial \tau_0} = I - \frac{\omega_0 J}{4\pi} - \left( 1 - \omega_0 \right) B,
\label{eq:rt_isotropic}
\end{equation}
where $B$ is the blackbody/Planck function.  The quantity $\omega_0$ is the ratio of the scattering cross section to the total (absorption and scattering) cross section and is termed the ``single-scattering albedo".  The thermal emission is assumed to be in local thermodynamic equilibrium (LTE).  By ``coherent", we mean that the incoming and outgoing photons have the same frequency.  Traditionally, such an approximation is used to describe the continuum in stellar atmospheres.  It is a bad approximation for spectral lines, unless they have zero width and the scattering atoms or molecules are completely at rest.  Instead, spectral lines are better described by the opposite limit of ``complete redistribution" (or ``complete noncoherence"), where the frequency of the outgoing photons are randomly redistributed over the line profile \citep{mihalas}.  In highly-irradiated exoplanets, coherent scattering is a decent approximation, because of the presence of dense forests of lines and collision-induced absorption, the latter of which functions like absorption by a continuum.  Redistribution over each individual line, in such a dense forest of lines, is then a relatively minor effect.  The problem is further alleviated if synthetic spectra are computed over relatively broad wavelength bins.

There are two ways to proceed.  The first is to solve equation (\ref{eq:rt_isotropic}) using $\tau_0$ as the independent variable (e.g., see Appendix \ref{append:direct} for the case of pure absorption).  The second is to relate $\tau_0$ to a slant optical depth ($\tau$) via some characteristic value of $\mu$ ($\bar{\mu}$).  \cite{pierrehumbert} has previously discussed the various choices of $\bar{\mu}$ and adopted $\bar{\mu}=1/2$.  As an example, we note that \cite{fhz06} also use $\bar{\mu}=1/2$.  In order to facilitate comparison with studies that generally make choices for the value of $\bar{\mu}$---sometimes without explicitly stating them---we will seek two-stream solutions with $\tau$ as the independent variable and leave the value of $\bar{\mu}$ unspecified.

To transform equation (\ref{eq:rt_isotropic}) into its two-stream form, we first rewrite it in terms of the slant optical depth,
\begin{equation}
\tau \equiv \frac{\tau_0}{\bar{\mu}_+},
\end{equation}
where $\bar{\mu}_+ \ge 0$ is a characteristic or mean value of $\mu$ in the upper hemisphere (defined by $0 \le \theta \le 90^\circ$ or $0 \le \mu \le 1$).  By integrating equation (\ref{eq:rt_isotropic}) over $\mu$ and using equations (\ref{eq:moments}) and (\ref{eq:moments2}), we obtain
\begin{equation}
\begin{split}
\frac{\partial F_\uparrow}{\partial \tau} =& \bar{\mu}_+ F_\uparrow \left[ \frac{1}{\epsilon_+} - \frac{\omega_0 }{2 \epsilon} \right] \\
&- \frac{\bar{\mu}_+ \omega_0}{2 \epsilon} F_\downarrow - 2 \pi \bar{\mu}_+ \left(1 - \omega_0 \right) B.
\end{split}
\label{eq:twostream_up_iso}
\end{equation}
In the lower hemisphere (defined by $90^\circ \le \theta \le 180^\circ$ or $-1 \le \mu \le 0$), we define
\begin{equation}
\tau \equiv \frac{\tau_0}{\bar{\mu}_-},
\end{equation}
where $\bar{\mu}_- \le 0$ is a characteristic or mean value of $\mu$, and then integrate equation (\ref{eq:rt_isotropic}) to obtain
\begin{equation}
\begin{split}
\frac{\partial F_\downarrow}{\partial \tau} =& \bar{\mu}_- F_\downarrow \left[ \frac{1}{\epsilon_-} - \frac{\omega_0 }{2 \epsilon} \right] \\
&- \frac{\bar{\mu}_- \omega_0}{2 \epsilon} F_\uparrow - 2 \pi \bar{\mu}_- \left(1 - \omega_0 \right) B.
\end{split}
\label{eq:twostream_down_iso}
\end{equation}

In transforming the radiative transfer equation, which involves the intensity, into its moments, which involve the total intensity, fluxes and other higher moments, one needs a series of ``closures", which effectively reduce the number of unknown variables by one---the number of unknowns now becomes equal to the number of equations.  These closures are generally termed the ``Eddington coefficients" \citep{mm99}, although there appears to be no consensus on how to number them.  In the present study, we will number the Eddington coefficients in the order in which we will invoke them.  In the case of isotropic, coherent scattering, we define the following ``first Eddington coefficients", 
\begin{equation}
\epsilon_+ \equiv \frac{F_\uparrow}{J_\uparrow}, ~\epsilon_- \equiv \frac{F_\downarrow}{J_\downarrow}, ~\epsilon \equiv \frac{F_+}{J},
\end{equation}
where there is one each for the outgoing/upper hemisphere ($\epsilon_+$), the incoming/lower hemisphere ($\epsilon_-$) and the entire atmosphere ($\epsilon$).  We will see later that the values of these first Eddington coefficients may be fixed via a series of physical constraints.

We assume that the Eddington coefficients in the outgoing and incoming hemispheres are equal, i.e., $\epsilon_\pm = \epsilon_+ = \epsilon_-$.  Furthermore, we assume that the characteristic values of $\mu$ have the same magnitude in each hemisphere,
\begin{equation}
\bar{\mu} = \bar{\mu}_+ = -\bar{\mu}_-.
\end{equation}
These assumptions are commonly made, but seldom explicitly elucidated (e.g., \citealt{pierrehumbert}).  Any physical process that leads to an asymmetry between the outgoing and incoming values of the flux and mean intensity will render $\epsilon_+ \ne \epsilon_-$ and $\bar{\mu}_+ \ne \bar{\mu}_-$.  Furthermore, $\epsilon_+$, $\epsilon_-$ and $\epsilon$ are generally not expected to be constant with pressure or height in an atmosphere.

With these assumptions, the pair of equations in (\ref{eq:twostream_up_iso}) and (\ref{eq:twostream_down_iso}) may be rewritten in a more compact form,
\begin{equation}
\begin{split}
\frac{\partial F_\uparrow}{\partial \tau} =& \gamma_{\rm a} F_\uparrow - \gamma_{\rm s} F_\downarrow - \gamma_{\rm B} B, \\
\frac{\partial F_\downarrow}{\partial \tau} =& - \gamma_{\rm a} F_\downarrow + \gamma_{\rm s} F_\uparrow + \gamma_{\rm B} B,
\end{split}
\label{eq:twostream_iso}
\end{equation}
where the coefficients of the equations are
\begin{equation}
\begin{split}
\gamma_{\rm a} &\equiv \bar{\mu} \left( \frac{1}{\epsilon_\pm} - \frac{\omega_0}{2 \epsilon} \right), \\
\gamma_{\rm s} &\equiv \frac{\bar{\mu} \omega_0}{2 \epsilon}, \\
\gamma_{\rm B} &\equiv 2 \pi \bar{\mu} \left( 1 - \omega_0 \right).
\end{split}
\label{eq:twostream_iso_coefficients_1}
\end{equation}
We note that the pair of equations in (\ref{eq:twostream_iso}) have the same mathematical form as equations (11) and (12) of \cite{toon89}.  Instead of using the generic labels of ``$\gamma_1$" and ``$\gamma_2$" for the coefficients, we have used $\gamma_{\rm a}$ and $\gamma_{\rm s}$ to refer to the effects of absorption (via the subscript ``a") and scattering (via the subscript ``s").  In a purely-absorbing atmosphere, we have $\gamma_{\rm s} = 0$, such that outgoing rays remain outgoing and incoming rays remain incoming, at least for the two-stream approximation.  Scattering converts some of the outgoing rays into incoming ones (and vice versa), because $\gamma_{\rm s} \ne 0$.

\cite{mw80} and \cite{toon89} have previously derived equivalent forms of equation (\ref{eq:twostream_iso}) from the radiative transfer equation, while \cite{pierrehumbert} has previously stated equation (\ref{eq:twostream_iso}) in a heuristic way (see his Chapter 5.5).

\subsection{Enforcing Energy Conservation in Purely Scattering Limit}
\label{subsect:enforce_scat}

Even without solving the pair of equations in (\ref{eq:twostream_iso}), we may simplify the expression for $\gamma_{\rm a}$ and $\gamma_{\rm s}$ by demanding that energy is conserved in the purely scattering limit ($\omega_0=1$, $\gamma_{\rm B}=0$), which yields
\begin{equation}
\frac{\partial F_-}{\partial \tau} = \left( \gamma_{\rm a} - \gamma_{\rm s} \right) F_+ = 0.
\end{equation}
Since we generally expect $F_+ \ne 0$, this implies that we must have $\gamma_{\rm a} = \gamma_{\rm s}$, which yields
\begin{equation}
\epsilon = \epsilon_\pm.
\end{equation}
It follows that
\begin{equation}
\begin{split}
\gamma_{\rm a} &= \frac{\bar{\mu}}{\epsilon_\pm} \left( 1 - \frac{\omega_0}{2} \right), \\
\gamma_{\rm s} &= \frac{\bar{\mu} \omega_0}{2 \epsilon_\pm}.
\end{split}
\end{equation}
This line of reasoning was previously employed by \cite{toon89}.

\subsection{Enforcing Correct Total Blackbody Flux in Isothermal, Opaque, Purely Absorbing Atmosphere}
\label{subsect:enforce_abs}

In the purely absorbing limit, we enforce the condition that the blackbody flux emitted by an opaque, isothermal\footnote{Specifically, we assume that $B$ is constant with $\tau$.} atmosphere is correct.  When $\omega_0=0$, we get $\gamma_{\rm a} = \bar{\mu}/\epsilon_\pm$, $\gamma_{\rm s}=0$, $\gamma_{\rm B} = 2 \pi \bar{\mu}$ and the solutions to the equations in (\ref{eq:twostream_iso}) are
\begin{equation}
\begin{split}
F_\uparrow &= F_{\uparrow_0} \exp{\left(\gamma_{\rm a} \tau \right)} + \frac{\gamma_{\rm B} B}{\gamma_{\rm a}} \left[ 1 - \exp{\left(\gamma_{\rm a} \tau \right)} \right], \\
F_\downarrow &= F_{\downarrow_0} \exp{\left(-\gamma_{\rm a} \tau \right)} + \frac{\gamma_{\rm B} B}{\gamma_{\rm a}} \left[ 1 - \exp{\left(-\gamma_{\rm a} \tau \right)} \right], \\
\end{split}
\label{eq:infinite_isothermal_solution}
\end{equation}
where $F_{\uparrow_0}$ and $F_{\downarrow_0}$ are the values of $F_\uparrow$ and $F_\downarrow$, respectively, when $\tau=0$.  In an opaque atmosphere, we have $F_\uparrow \rightarrow \gamma_{\rm B} B/\gamma_{\rm a}$ as $\tau \rightarrow -\infty$, while $F_\downarrow \rightarrow \gamma_{\rm B} B/\gamma_{\rm a}$ as $\tau \rightarrow \infty$.  This implies that the total flux becomes
\begin{equation}
F_+ \rightarrow \frac{2\gamma_{\rm B} B}{\gamma_{\rm a}} = 4 \pi \epsilon_\pm B.
\end{equation}
By assuming the blackbody radiation to be isotropic over each hemisphere, one may show that
\begin{equation}
\int^{2\pi}_0 \int^1_0 \mu B d\mu d\phi = \int^{2\pi}_0 \int^0_{-1} \mu B d\mu d\phi = \pi B.
\end{equation}
Since the correct limit is $F_+ \rightarrow 2 \pi B$, this implies that $\epsilon_\pm = 1/2$.  It follows that
\begin{equation}
\begin{split}
\gamma_{\rm a} &= \bar{\mu} \left( 2 - \omega_0 \right), \\
\gamma_{\rm s} &= \bar{\mu} \omega_0.
\end{split}
\label{eq:twostream_iso_coefficients_2}
\end{equation}
The same line of reasoning was again previously employed by \cite{toon89}.  Within our formalism, $1/\epsilon_\pm$ is the diffusivity factor (see \S\ref{subsect:diffusivity}).

\subsection{Equivalence of Solving First- and Second-Order Differential Equations}
\label{subsect:equivalence}

In equation (\ref{eq:infinite_isothermal_solution}), we previously stated the solution for an isothermal slab bounded by 0 and $\tau$.  There are two approaches to solving the two-stream equations: either as a pair of first-order differential equations or as a single second-order differential equation.  Identical answers are obtained if the correct boundary conditions are specified.  The equivalence of these approaches may be cleanly demonstrated using the pair of two-stream equations in the purely absorbing limit,
\begin{equation}
\begin{split}
\frac{\partial F_\uparrow}{\partial \tau} =& 2 \bar{\mu} F_\uparrow - 2 \pi \bar{\mu} B, \\
\frac{\partial F_\downarrow}{\partial \tau} =& - 2\bar{\mu} F_\downarrow + 2\pi \bar{\mu} B.
\end{split}
\label{eq:twostream_iso_abs}
\end{equation}
While the mathematical techniques presented in this subsection are well-known and not novel, we review them within the context of our problem so that we may apply them later in \S\ref{subsect:general_solution_iso}, \S\ref{subsect:non_iso_non_isothermal} and Appendix \ref{append:convective}.

Solving the pair of equations in (\ref{eq:twostream_iso_abs}) involves realizing that
\begin{equation}
\begin{split}
\frac{\partial}{\partial \tau} \left[ F_\uparrow \exp{\left( -2\bar{\mu} \tau \right)} \right] &= \exp{\left( -2\bar{\mu} \tau \right)} \frac{\partial F_\uparrow}{\partial \tau} - 2 \bar{\mu} F_\uparrow \exp{\left( -2\bar{\mu} \tau \right)}, \\
\frac{\partial}{\partial \tau} \left[ F_\downarrow \exp{\left( 2\bar{\mu} \tau \right)} \right] &= \exp{\left( 2\bar{\mu} \tau \right)} \frac{\partial F_\downarrow}{\partial \tau} + 2 \bar{\mu} F_\downarrow \exp{\left( 2\bar{\mu} \tau \right)}. \\
\end{split}
\end{equation}
Integrating between two layers, with optical depths of $\tau_1$ and $\tau_2$ (where $\tau_2 > \tau_1$), we obtain
\begin{equation}
\begin{split}
F_{\uparrow_1} =& F_{\uparrow_2} \exp{\left[ 2\bar{\mu} \left(\tau_1-\tau_2\right) \right]} \\
&+ 2\pi \bar{\mu} \int^{\tau_2}_{\tau_1} B \exp{\left[ 2\bar{\mu} \left(\tau_1-\tau \right) \right]} d\tau, \\
F_{\downarrow_2} =& F_{\downarrow_1} \exp{\left[ 2\bar{\mu} \left(\tau_1-\tau_2\right) \right]} \\
&+ 2\pi \bar{\mu} \int^{\tau_2}_{\tau_1} B \exp{\left[ 2\bar{\mu} \left(\tau-\tau_2 \right) \right]} d\tau.
\end{split}
\end{equation}
We have intentionally written the expression for the outgoing flux in this way, because it is obtained by integrating upwards from the boundary condition at the bottom of the atmosphere (BOA).  In this manner, $F_\uparrow$ is computed for the layer immediately above the BOA.  This procedure is repeated until each of the model atmospheric layers has a computed value of $F_\uparrow$.  Similarly, the incoming flux is obtained by integrating downwards from the boundary condition at the top of the atmosphere (TOA) and populating each layer with a computed value of $F_\downarrow$.  In the isothermal limit, we obtain
\begin{equation}
\begin{split}
F_{\uparrow_1} =& F_{\uparrow_2} \exp{\left[ 2\bar{\mu} \left(\tau_1-\tau_2\right) \right]} + \pi B \left\{ 1 - \exp{\left[ 2\bar{\mu} \left(\tau_1-\tau_2\right) \right]} \right\}, \\
F_{\downarrow_2} =& F_{\downarrow_1} \exp{\left[ 2\bar{\mu} \left(\tau_1-\tau_2\right) \right]} + \pi B \left\{ 1 - \exp{\left[ 2\bar{\mu} \left(\tau_1-\tau_2\right) \right]} \right\}. \\
\end{split}
\label{eq:twostream_iso_abs_solutions}
\end{equation}

The approach of solving a pair of first-order differential equations becomes challenging when the equations for $F_\uparrow$ and $F_\downarrow$ are coupled in the presence of scattering.  A more general approach is to cast the problem in terms of a second-order differential equation for $F_+$, which is applicable even when $\gamma_{\rm s} \ne 0$.  By separately adding and subtracting the pair of equations in (\ref{eq:twostream_iso_abs}), we get
\begin{equation}
\begin{split}
\frac{\partial F_+}{\partial \tau} =& 2 \bar{\mu} F_-, \\
\frac{\partial F_-}{\partial \tau} =& 2\bar{\mu} F_+ - 4\pi \bar{\mu} B,
\end{split}
\label{eq:twostream_iso_abs_2}
\end{equation}
from which it follows that
\begin{equation}
\frac{\partial^2 F_+}{\partial \tau^2} - 4\bar{\mu}^2 F_+ = - 8\pi \bar{\mu}^2 B.
\label{eq:twostream_iso_abs_3}
\end{equation}
This second-order differential equation has the solution,
\begin{equation}
F_+ = {\cal A}_1 \exp{\left( 2 \bar{\mu} \tau \right)} + {\cal A}_2 \exp{\left( -2 \bar{\mu} \tau \right)} + 2 \pi B.
\end{equation}
The coefficients ${\cal A}_1$ and ${\cal A}_2$ are determined by imposing a pair of boundary conditions.  To keep the algebra tractable for now and merely illustrate the method, we have assumed isothermality ($\partial B/\partial \tau = 0$) for this subsection.  It follows that
\begin{equation}
F_- = {\cal A}_1 \exp{\left( 2 \bar{\mu} \tau \right)} - {\cal A}_2 \exp{\left( -2 \bar{\mu} \tau \right)}.
\end{equation}
By again imposing the boundary conditions $F_{\uparrow_2}$ and $F_{\downarrow_1}$, we obtain
\begin{equation}
\begin{split}
{\cal A}_1 &= \left( F_{\uparrow_2} - \pi B \right) \exp{\left( -2 \bar{\mu} \tau_2 \right)}, \\
{\cal A}_2 &= \left( F_{\downarrow_1} - \pi B \right) \exp{\left( 2 \bar{\mu} \tau_1 \right)}, \\
\end{split}
\end{equation}
from which we may derive the pair of equations in (\ref{eq:twostream_iso_abs_solutions}).

Thus, the two approaches of either solving a pair of first-order differential equations (for $F_\uparrow$ and $F_\downarrow$) or a single second-order differential equation (for $F_+$) are equivalent, at least in the purely absorbing, isothermal limit.  In more general cases, we will use the method of solving the second-order differential equation.

\subsection{General Solution with Isotropic Scattering and Non-Isothermal Layers}
\label{subsect:general_solution_iso}

We now return to solving equation (\ref{eq:twostream_iso}) in the general sense.  Adding and substracting the equations in turn yields,
\begin{equation}
\begin{split}
\frac{\partial F_+}{\partial \tau} =& 2 \bar{\mu} F_-, \\
\frac{\partial F_-}{\partial \tau} =& 2\bar{\mu} \left( 1 - \omega_0 \right) \left( F_+ - 2\pi B \right),
\end{split}
\label{eq:twostream_iso_2}
\end{equation}
from which we obtain
\begin{equation}
\frac{\partial^2 F_+}{\partial \tau^2} - \alpha^2 F_+ = - 2\pi \alpha^2 B,
\label{eq:twostream_iso_3}
\end{equation}
where we have defined
\begin{equation}
\alpha \equiv 2 \bar{\mu} \left( 1 - \omega_0 \right)^{1/2}.
\label{eq:alpha}
\end{equation}
The homogeneous solution to equation (\ref{eq:twostream_iso_3}) is as before,
\begin{equation}
F_{+{\rm h}} = {\cal A}_1 \exp{\left( \alpha \tau \right)} + {\cal A}_2 \exp{\left( -\alpha \tau \right)}.
\end{equation}

The particular solution depends on the functional form adopted for $B$.  Generally, we expect each model layer to have an \textit{internal} temperature gradient, implying that $B$ depends on $\tau$, since $T$ depends on $\tau$.  Following \cite{toon89}, we write $B$ as a linear function of $\tau$,
\begin{equation}
B = B_0 + B^\prime \bar{\mu} \left( \tau - \tau^\prime \right),
\label{eq:b_expand}
\end{equation} 
where 
\begin{equation}
B^\prime \equiv \frac{1}{\bar{\mu}} \frac{\partial B}{\partial \tau} \approx \frac{B_2 - B_1}{ \bar{\mu} \left( \tau_2 - \tau_1 \right)}
\end{equation}
is the gradient of the Planck function across a given layer and is constant for that layer.  The quantity $\tau^\prime$ is present in equation (\ref{eq:b_expand}) to translate the ``zero" of the optical depth to the edge of a layer.\footnote{A potential source of confusion comes from the fact that \cite{toon89} use two contradicting definitions for their $\tau$.  In the text below their equation (1), it is described as the optical depth measured from the top of the atmosphere.  In their equation (25), it is the optical depth measured from the edge of a model layer.}  The quantities $B_1$ and $B_2$ are the Planck function evaluated at $\tau_1$ and $\tau_2$, respectively.  The values of the quantities $B_0$ and $\tau^\prime$ depend on whether one is dealing with the outgoing or incoming flux.  For the outgoing flux, $B_0 = B_2$ and $\tau^\prime = \tau_2$.  For the incoming flux, $B_0 = B_1$ and $\tau^\prime = \tau_1$.  These choices ensure that when $\tau=\tau_{1,2}$, equation (\ref{eq:b_expand}) gives $B=B_{1,2}$.  Mathematically, equation (\ref{eq:b_expand}) qualifies as the Taylor series expansion of the Planck function about the point $\tau=\tau^\prime$, truncated at the linear term.

With this choice of $B$, the particular solution takes the form,
\begin{equation}
F_{+{\rm p}} = 2\pi B.
\end{equation}
The full solution (homogeneous plus particular) to equation (\ref{eq:twostream_iso_3}) is
\begin{equation}
F_+ = {\cal A}_1 \exp{\left( \alpha \tau \right)} + {\cal A}_2 \exp{\left( -\alpha \tau \right)} + 2 \pi B,
\label{eq:fplus}
\end{equation}
from which we obtain
\begin{equation}
F_- = \frac{\alpha}{2 \bar{\mu}} \left[ {\cal A}_1 \exp{\left( \alpha \tau \right)} - {\cal A}_2 \exp{\left( -\alpha \tau \right)} \right] + \pi B^\prime.
\label{eq:fminus}
\end{equation}
The outgoing and incoming fluxes are
\begin{equation}
\begin{split}
F_{\uparrow} =& {\cal A}_1 \zeta_+ \exp{\left( \alpha \tau \right)} + {\cal A}_2 \zeta_- \exp{\left( -\alpha \tau \right)} + \pi B + \frac{\pi B^\prime}{2}, \\
F_{\downarrow} =& {\cal A}_1 \zeta_- \exp{\left( \alpha \tau \right)} + {\cal A}_2 \zeta_+ \exp{\left( -\alpha \tau \right)} + \pi B - \frac{\pi B^\prime}{2}, \\
\end{split}
\end{equation}
where we have defined
\begin{equation}
\zeta_\pm \equiv \frac{1}{2} \left[ 1 \pm \left( 1 - \omega_0 \right)^{1/2} \right].
\label{eq:zeta}
\end{equation}

To derive expressions for the coefficients ${\cal A}_1$ and ${\cal A}_2$, we have to impose the boundary conditions $F_{\uparrow_2}$ and $F_{\downarrow_1}$, which yields
\begin{equation}
\begin{split}
F_{\uparrow_2} =& {\cal A}_1 \zeta_+ \exp{\left( \alpha \tau_2 \right)} + {\cal A}_2 \zeta_- \exp{\left( -\alpha \tau_2 \right)} + \pi {\cal B}_{2+}, \\
F_{\downarrow_1} =& {\cal A}_1 \zeta_- \exp{\left( \alpha \tau_1 \right)} + {\cal A}_2 \zeta_+ \exp{\left( -\alpha \tau_1 \right)} + \pi {\cal B}_{1-} , \\
\end{split}
\label{eq:bc_iso}
\end{equation}
where we have found it convenient to define the quantities,
\begin{equation}
\begin{split}
{\cal B}_{i-} &\equiv B_1 + B^\prime \bar{\mu} \left( \tau_i - \tau_1 \right) - \frac{B^\prime}{2}, \\
{\cal B}_{i+} &\equiv B_2 + B^\prime \bar{\mu} \left( \tau_i - \tau_2 \right) + \frac{B^\prime}{2}. \\
\end{split}
\end{equation}

A more intuitive way of writing down the solutions for the outgoing and incoming fluxes is to cast them in terms of the transmission function (or simply the ``transmissivity") \citep{pierrehumbert},
\begin{equation}
{\cal T} \equiv \exp{\left[ -\alpha \left( \tau_2 - \tau_1 \right) \right]},
\end{equation}
noting that $\tau_2 > \tau_1$.  This approach is also more ideal for computation, since we have $0 \le {\cal T} \le 1$ (instead of unwieldy exponentials with potentially large exponents).  The task is to derive expressions for $F_{\uparrow_1}$ in terms of $F_{\uparrow_2}$ and ${\cal T}$, and also $F_{\downarrow_2}$ in terms of $F_{\downarrow_1}$ and ${\cal T}$.  More specifically, we have to find expressions for ${\cal A}_1 \zeta_+ \exp{(\alpha \tau_1)}$ and ${\cal A}_2 \zeta_- \exp{(-\alpha \tau_1)}$ when deriving $F_{\uparrow_1}$.  For $F_{\downarrow_2}$, we need expressions for ${\cal A}_1 \zeta_- \exp{(\alpha \tau_2)}$ and ${\cal A}_2 \zeta_+ \exp{(-\alpha \tau_2)}$.  Manipulating the pair of expressions in (\ref{eq:bc_iso}) gives
\begin{equation}
\begin{split}
&{\cal A}_2 \exp{\left( -\alpha \tau_1 \right)} = \frac{1}{\zeta_-^2 {\cal T} - \zeta^2_+ {\cal T}^{-1}} \left[ \zeta_- F_{\uparrow_2} - \zeta_+ {\cal T}^{-1} F_{\downarrow_1} \right.\\
&- \left. \pi \left( \zeta_- {\cal B}_{2+} - \zeta_+ {\cal T}^{-1} {\cal B}_{1-} \right)  \right],
\end{split}
\end{equation}
and 
\begin{equation}
\begin{split}
&{\cal A}_2 \exp{\left( -\alpha \tau_2 \right)} = \frac{1}{\zeta_-^2 {\cal T} - \zeta^2_+ {\cal T}^{-1}} \left[ \zeta_- {\cal T} F_{\uparrow_2} - \zeta_+ F_{\downarrow_1} \right.\\
&- \left. \pi \left( \zeta_- {\cal T} {\cal B}_{2+} - \zeta_+ {\cal B}_{1-} \right) \right].
\end{split}
\end{equation}

The expressions in (\ref{eq:bc_iso}) permit two ways of deriving ${\cal A}_1 \zeta_+ \exp{(\alpha \tau_1)}$ or ${\cal A}_1 \zeta_- \exp{(\alpha \tau_2)}$.  One can choose to use either the equation involving the boundary condition $F_{\uparrow_2}$ or $F_{\downarrow_1}$.  In deriving $F_{\uparrow_1}$, we use the first equation in (\ref{eq:bc_iso}).  In deriving $F_{\downarrow_2}$, we use the second equation in (\ref{eq:bc_iso}).  It follows that
\begin{equation}
\begin{split}
{\cal A}_1 \zeta_+ \exp{\left( \alpha \tau_1 \right)} =& {\cal T} F_{\uparrow_2} - {\cal A}_2 \zeta_- {\cal T}^2 \exp{\left( - \alpha \tau_1 \right)} \\
&- \pi {\cal B}_{2+} {\cal T}, \\
{\cal A}_1 \zeta_- \exp{\left( \alpha \tau_2 \right)} =& {\cal T}^{-1} F_{\downarrow_1} - {\cal A}_2 \zeta_+ {\cal T}^{-2} \exp{\left( - \alpha \tau_2 \right)} \\
&- \pi {\cal B}_{1-} {\cal T}^{-1}. \\
\end{split}
\end{equation}

Assembling all of the various pieces enables us to obtain
\begin{equation}
\begin{split}
F_{\uparrow_1} =& \frac{1}{\left( \zeta_- {\cal T} \right)^2 - \zeta_+^2} \left\{ \left( \zeta_-^2 - \zeta_+^2 \right) {\cal T} F_{\uparrow_2} - \zeta_- \zeta_+ \left( 1 - {\cal T}^2 \right) F_{\downarrow_1} \right. \\
&+ \left. \pi \left[ {\cal B}_{1+} \left( \zeta^2_- {\cal T}^2 - \zeta_+^2 \right) + {\cal B}_{2+} {\cal T}  \left( \zeta_+^2 - \zeta_-^2 \right) \right. \right. \\
&+ \left. \left. {\cal B}_{1-} \zeta_- \zeta_+ \left( 1 - {\cal T}^2 \right) \right] \right\},  \\
F_{\downarrow_2} =& \frac{1}{\left( \zeta_- {\cal T} \right)^2 - \zeta_+^2} \left\{ \left( \zeta_-^2 - \zeta_+^2 \right) {\cal T} F_{\downarrow_1} - \zeta_- \zeta_+ \left( 1 - {\cal T}^2 \right) F_{\uparrow_2} \right. \\
&+ \left. \pi \left[ {\cal B}_{2-} \left( \zeta^2_- {\cal T}^2 - \zeta_+^2 \right) + {\cal B}_{1-} {\cal T}  \left( \zeta_+^2 - \zeta_-^2 \right) \right. \right. \\
&+ \left. \left. {\cal B}_{2+} \zeta_- \zeta_+ \left( 1 - {\cal T}^2 \right) \right] \right\}.  \\
\end{split}
\label{eq:twostream_iso_general_solution}
\end{equation}

In the limit of pure absorption ($\omega_0=0$), we have $\zeta_-=0$ and $\zeta_+=1$ and the two-stream solutions reduce to
\begin{equation}
\begin{split}
F_{\uparrow_1} =& {\cal T} F_{\uparrow_2} + \pi \left( {\cal B}_{1+} - {\cal B}_{2+} {\cal T} \right), \\
F_{\downarrow_2} =& {\cal T} F_{\downarrow_1} + \pi \left( {\cal B}_{2-} - {\cal B}_{1-} {\cal T} \right). \\
\end{split}
\label{eq:twostream_iso_pure_abs}
\end{equation}
In the isothermal limit, we recover equation (\ref{eq:twostream_iso_abs_solutions}).

Unlike in the purely absorbing limit, verifying the two-stream solutions in the limit of pure scattering is a subtler issue.  As already noted by \cite{toon89}, the two-stream solutions derived for $\omega_0 \ne 1$ are not valid in the limiting case of $\omega_0=1$.  One needs to return to the governing equations in (\ref{eq:twostream_iso_2}) in the limit of $\omega_0=1$ and solve them directly \citep{toon89}.  Specifically, the equations in (\ref{eq:twostream_iso_general_solution}) need to be replaced by
\begin{equation}
\begin{split}
F_{\uparrow_1} &= F_{\uparrow_2} - \frac{\left( F_{\uparrow_2} - F_{\downarrow_1} \right) \bar{\mu} \left( \tau_2 - \tau_1 \right)}{1 + \bar{\mu} \left( \tau_2 - \tau_1 \right)}, \\
F_{\downarrow_2} &= F_{\downarrow_1} + \frac{\left( F_{\uparrow_2} - F_{\downarrow_1} \right) \bar{\mu} \left( \tau_2 - \tau_1 \right)}{1 + \bar{\mu} \left( \tau_2 - \tau_1 \right)}, \\
\end{split}
\label{eq:twostream_iso_pure_scat}
\end{equation}
One may verify that when the model layer is opaque ($\tau_2-\tau_1 \gg 1/\bar{\mu}$), one recovers the pure reflection of the boundary conditions: $F_{\uparrow_1} = F_{\downarrow_1}$ and $F_{\downarrow_2} = F_{\uparrow_2}$.  When the layer is transparent ($\tau_2-\tau_1 \ll 1/\bar{\mu}$), we get $F_{\uparrow_1} = F_{\uparrow_2}$ and $F_{\downarrow_2} = F_{\downarrow_1}$.  

We note that when $\omega_0=1$, the equations in  (\ref{eq:twostream_iso_general_solution}) reduce to $F_{\uparrow_1} = F_{\downarrow_1} + \pi B^\prime$ and $F_{\downarrow_2} = F_{\uparrow_2} - \pi B^\prime$.  It almost reproduces the pure scattering limit for an opaque atmosphere, but with blackbody terms that produce unphysical contributions.

\subsection{General Solution with Isotropic Scattering and Isothermal Layers}

We now study trends in the limit of each model atmospheric layer being isothermal, where the two-stream solutions are
\begin{equation}
\begin{split}
F_{\uparrow_1} =& \frac{1}{\left( \zeta_- {\cal T} \right)^2 - \zeta_+^2} \left\{ \left( \zeta_-^2 - \zeta_+^2 \right) {\cal T} F_{\uparrow_2} - \zeta_- \zeta_+ \left( 1 - {\cal T}^2 \right) F_{\downarrow_1} \right. \\
&+ \left. \pi B \left[ \zeta_- \zeta_+ \left( 1 - {\cal T}^2 \right) - \left( \zeta_-^2 {\cal T} + \zeta_+^2 \right) \left( 1 - {\cal T} \right) \right] \right\},  \\
F_{\downarrow_2} =& \frac{1}{\left( \zeta_- {\cal T} \right)^2 - \zeta_+^2} \left\{ \left( \zeta_-^2 - \zeta_+^2 \right) {\cal T} F_{\downarrow_1} - \zeta_- \zeta_+ \left( 1 - {\cal T}^2 \right) F_{\uparrow_2} \right. \\
&+ \left. \pi B \left[ \zeta_- \zeta_+ \left( 1 - {\cal T}^2 \right) - \left( \zeta_-^2 {\cal T} + \zeta_+^2 \right) \left( 1 - {\cal T} \right) \right] \right\}.  \\
\end{split}
\label{eq:twostream_iso_general_solution_isothermal}
\end{equation}
Written in this form, the purpose of the ``coupling coefficients" becomes clear: $\zeta_\pm$ are order-of-unity, dimensionless coefficients that assign relative weights to the bottom and top boundary conditions, depending on the strength of scattering.  In the limits of pure absorption ($\zeta_-=0, \zeta_+=1$) or pure scattering ($\zeta_\pm=1/2$), this coupling is broken.  In between, it depends on the symmetry properties of scattering (Figure \ref{fig:zetas}).  The ratio $\zeta_-/\zeta_+$ is also the spherical albedo, as we will see in \S\ref{subsect:shortwave_albedo}.  Generalized expressions for $\zeta_\pm$, involving non-isotropic scattering, will be derived in \S\ref{sect:non_isotropic}.

When the layers are transparent (${\cal T}=1$), we have $F_{\uparrow_1} = F_{\uparrow_2}$ and $F_{\downarrow_2} = F_{\downarrow_1}$, as expected.  In opaque layers (${\cal T}=0$), we have
\begin{equation}
\begin{split}
F_{\uparrow_1} =& \frac{ \zeta_- F_{\downarrow_1}}{\zeta_+} + \pi B \left( 1 - \frac{\zeta_-}{\zeta_+} \right), \\
F_{\downarrow_2} =& \frac{ \zeta_- F_{\uparrow_2}}{\zeta_+} + \pi B \left( 1 - \frac{\zeta_-}{\zeta_+} \right). \\
\end{split}
\end{equation}
The factor $\zeta_-/\zeta_+$ is a steep function of $\omega_0$ (Figure \ref{fig:zetas}), implying that the fluxes rapidly converge towards the boundary conditions as scattering becomes more dominant.

It is apparent from equation (\ref{eq:twostream_iso_general_solution_isothermal}) that if the boundary conditions $F_{\uparrow_2}$ and $F_{\downarrow_1}$ assume equal values, then the outgoing and incoming fluxes are identical.  Thus, to illustrate the diversity of solutions possible, we adopt $F_{\uparrow_2}/\pi B = 0$ and $F_{\downarrow_1}/\pi B = 1$.  Figure \ref{fig:fluxes_iso} illustrates several basic trends.  As expected, we get $F_{\uparrow_1} \rightarrow F_{\downarrow_1}$ and $F_{\downarrow_2} \rightarrow F_{\uparrow_2}$ as $\omega_0 \rightarrow 1$, independent of ${\cal T}$, i.e., pure reflection of the boundary conditions.  The incoming flux ($F_{\downarrow_2}$) increases with the transmission (${\cal T}$) as one expects for an atmosphere that is irradiated from above.  Curiously, the outgoing flux ($F_{\uparrow_1}$) \textit{decreases} as the transmission increases, but this is a consequence of the fact that there is no internal heat specified ($F_{\uparrow_2}=0$); it tends towards this vanishing boundary condition as the transmission increases.

We note that there is no contradiction between specifying ${\cal T}$ and $\omega_0$ as independent parameters.  A largely transparent atmospheric layer (${\cal T} \sim 1$) may still be purely absorbing ($\omega_0=0$) or scattering ($\omega_0=1$)---it just does not absorb or scatter enough to render itself opaque to radiation.  The transmission specifies the fraction of radiation passing through a layer, while the single-scattering albedo describes the relative strength of scattering versus absorption.

\begin{figure}
\begin{center}
\includegraphics[width=\columnwidth]{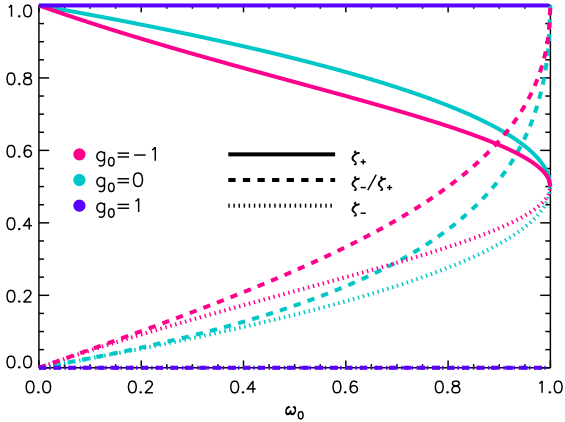}
\end{center}
\vspace{-0.2in}
\caption{Behavior of $\zeta_-$, $\zeta_+$ and their ratio as a function of the single-scattering albedo ($\omega_0$) for different values of the scattering asymmetry factor ($g_0$).  Note that $\zeta_-/\zeta_+$ is also the spherical albedo.}
\label{fig:zetas}
\end{figure}

\begin{figure}
\begin{center}
\includegraphics[width=\columnwidth]{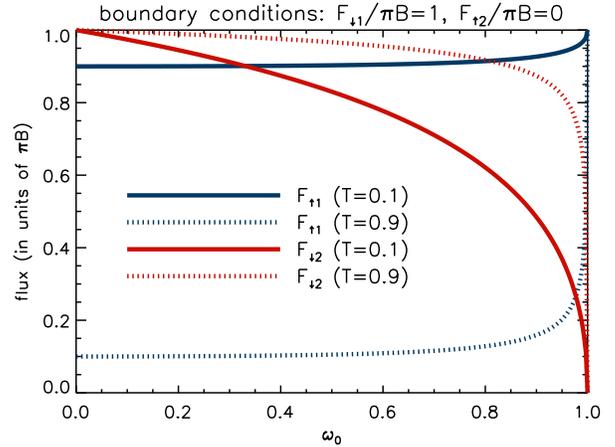}
\end{center}
\vspace{-0.2in}
\caption{Outgoing and incoming fluxes as functions of the single-scattering albedo ($\omega_0$) and for various values of the transmission (${\cal T}$) in the limit of isotropic, coherent scattering.  All fluxes and their boundary conditions are given in terms of the blackbody flux ($\pi B$).}
\vspace{0.1in}
\label{fig:fluxes_iso}
\end{figure}

\section{Two-Stream Radiative Transfer: Non-Isotropic, Coherent Scattering}
\label{sect:non_isotropic}

Naturally, the next generalization is to allow for coherent but non-isotropic scattering within the model atmosphere.  The governing equation now reads \citep{chandra60,mihalas,gy89},
\begin{equation}
\mu \frac{\partial I}{\partial \tau_0} = I - \frac{\omega_0}{4\pi} \int^{4\pi}_0 {\cal P} I d\Omega^\prime - \left( 1 - \omega_0 \right) B,
\label{eq:rt_non_isotropic}
\end{equation}
where ${\cal P}$ is the scattering phase function.  It is integrated over all incident angles in spherical coordinates $(\theta^\prime, \phi^\prime)$ such that $d\Omega^\prime \equiv d \mu^\prime d\phi^\prime$, where we have defined $\mu^\prime \equiv \cos\theta^\prime$.  Note that if ${\cal P}=1$, then we recover equation (\ref{eq:rt_isotropic}).

\subsection{General Properties of the Scattering Phase Function}

We can state a few general properties of the scattering phase function that will allow us to transform equation (\ref{eq:rt_non_isotropic}) into its two-stream form, even without explicitly specifying the functional form of ${\cal P}$.  Our derivation fills in details previously left out by other works.

\subsubsection{Sole Dependence on the Relative Scattering Angle}
\label{subsect:angle_symmetry}

Generally, it is assumed that ${\cal P} = {\cal P}(\Theta)$ only, where $\Theta = \theta^\prime - \theta$.  This assumption alone allows one to derive a symmetry property associated with $\mu^\prime$ and $\mu$ \citep{gy89,pierrehumbert}.  Consider two arbitrary locations in the atmosphere represented by the vectors $\vec{r^\prime} = (r^\prime \sin\theta^\prime \cos\phi^\prime, r^\prime \sin\theta^\prime \sin\phi^\prime, r^\prime \cos\theta^\prime)$ and $\vec{r} = (r \sin\theta \cos\phi, r \sin\theta \sin\phi, r \cos\theta)$.  Taking their dot product yields an expression for $\mu^{\prime\prime} \equiv \cos\Theta$,
\begin{equation}
\mu^{\prime\prime} = \mu^\prime \mu + \left( 1 - \mu^{\prime2} \right)^{1/2} \left( 1 - \mu^2 \right)^{1/2} \cos\left(\phi^\prime - \phi \right),
\label{eq:Theta_symmetry}
\end{equation}
previously stated in equation (8.2) of \cite{gy89} and equation (5.8) of \cite{pierrehumbert}.  Equation (\ref{eq:Theta_symmetry}) informs us that $\mu^{\prime\prime}$ is invariant to double sign flips in $\mu^\prime$ and $\mu$, 
\begin{equation}
\mu^{\prime\prime} \rightarrow \mu^{\prime\prime} \mbox{ if } \mu^\prime \rightarrow -\mu^\prime \mbox{ and } \mu \rightarrow - \mu.
\label{eq:Theta_symmetry_2}
\end{equation}
Since $\mu^{\prime\prime}$ is single-valued for $0 \le \Theta \le 180^\circ$, we may conclude that $\Theta$ and thus ${\cal P}$ are invariant under this transformation.

\subsubsection{Normalization Symmetry}
\label{subsect:norm_symmetry}

It is assumed that integrating ${\cal P}$ over all angles yields the same normalization, regardless of the coordinate system the integration is performed in,
\begin{equation}
\int^{4\pi}_0 {\cal P} d\Omega = \int^{4\pi}_0 {\cal P} d\Omega^\prime = \int^{4\pi}_0 {\cal P} d\Omega^{\prime\prime} = 4 \pi,
\end{equation}
where we have defined $d\Omega^{\prime\prime} \equiv d\mu^{\prime\prime} d\phi^{\prime\prime}$ and $\phi^{\prime\prime} \equiv \phi^\prime - \phi$.  This property implies that we can always replace ${\cal P}$ by some function ${\cal P}^\prime={\cal P}^\prime(\Theta)$ and still perform the integration in any of the coordinate systems.

\subsection{Obtaining the Two-Stream Equations and Solutions}

With the properties of ${\cal P}$ stated, we can evaluate the moments of equation (\ref{eq:rt_non_isotropic}).  Following \S5.2 of \cite{pierrehumbert}, we multiply equation (\ref{eq:rt_non_isotropic}) by a function ${\cal H} = {\cal H}(\theta)$ and integrate over all angles $\theta$ and $\phi$,
\begin{equation}
\begin{split}
&\frac{1}{\bar{\mu}} \frac{\partial}{\partial \tau} \int^{2\pi}_0 \left( \int^1_0 \mu {\cal H} I d\mu - \int^0_{-1} \mu {\cal H} I d\mu \right) d\phi \\
&= \int^{4\pi}_0 {\cal H} I d\Omega - {\cal I} - \left( 1 - \omega_0 \right) \int^{4\pi}_0 {\cal H} B d\Omega,
\end{split}
\label{eq:rt_hform}
\end{equation}
where we have defined
\begin{equation}
\begin{split}
{\cal I} &\equiv \omega_0 \int^{4\pi}_0 {\cal G} I d\Omega^\prime, \\
{\cal G} &\equiv \frac{1}{4\pi} \int^{4\pi}_0 {\cal H} {\cal P} d\Omega.
\end{split}
\end{equation}
Note that the minus sign in the integral involving $\bar{\mu}$ in equation (\ref{eq:rt_hform}) comes from the characteristic value of $\mu$ being positive and negative in the outgoing and incoming hemispheres, respectively ($\bar{\mu} = \bar{\mu}_+ = -\bar{\mu}_- > 0$).

To evaluate ${\cal G}$ and ${\cal I}$, one has to specify the functional form of ${\cal H}$.  As already explained by \cite{pierrehumbert}, different choices of ${\cal H}$ will lead to different forms of the two-stream equations, i.e., with different expressions for the coefficients $\gamma_{\rm a}$ and $\gamma_{\rm s}$.

\subsubsection{Evaluating the Integrals ${\cal G}$ and ${\cal I}$}

When ${\cal H}=1$, we obtain ${\cal G}=1$, ${\cal I}=\omega_0 J$ and
\begin{equation}
\frac{\partial F_-}{\partial \tau} = \bar{\mu} \left( 1 - \omega_0 \right) \left( J - 4\pi B \right).
\label{eq:rt_hform2}
\end{equation}

The next natural choice is ${\cal H}=\mu$, because it allows us to introduce another Eddington coefficient into the formalism.  Several steps are involved in evaluating ${\cal G}$ and ${\cal I}$.  First, we write $\mu = \cos(\theta^\prime - \Theta)$, use the trigonometric angle subtraction rule and obtain
\begin{equation}
{\cal G} = \frac{1}{4\pi} \int^{4\pi}_0 \left[ \mu^\prime \mu^{\prime\prime} + \left( 1 - \mu^{\prime2} \right)^{1/2} \left( 1 - \mu^{\prime\prime2} \right)^{1/2} \right] {\cal P} ~d\Omega.
\end{equation}
Exploiting the property that ${\cal P}$ and $\mu^{\prime\prime}$ are invariant when $\mu^\prime \rightarrow -\mu^\prime$ and $\mu \rightarrow -\mu$ (see \S\ref{subsect:angle_symmetry}), we see that the first and second terms in ${\cal G}$ are even and odd integrals, respectively.  The second term vanishes.  Second, we use the normalization symmetry described in \S\ref{subsect:norm_symmetry} to further write ${\cal G}$ as
\begin{equation}
{\cal G} = \frac{1}{4\pi} \int^{4\pi}_0 \mu^\prime \mu^{\prime\prime} {\cal P} d\Omega^{\prime\prime}.
\label{eq:gfunc}
\end{equation}
Note that this step is valid only because the integrand in equation (\ref{eq:gfunc}) does not depend on $\mu$ and is being evaluated at a \textit{fixed} value of $\mu^\prime$, as ${\cal G}$ is part of the integrand of ${\cal I}$.  Third, if we define the asymmetry factor as \citep{gy89,pierrehumbert}
\begin{equation}
g_0 \equiv \frac{1}{4\pi} \int^{4\pi}_0 \mu^{\prime\prime} {\cal P} d\Omega^{\prime\prime},
\label{eq:asym_factor}
\end{equation}
then we obtain the result,
\begin{equation}
{\cal G} = g_0 \mu^\prime,
\end{equation}
which was previously stated, without proof, in equation (8.142) of \cite{gy89} and equation (5.18) of \cite{pierrehumbert}.  It implies that
\begin{equation}
{\cal I} = \omega_0 g_0 F_+.
\end{equation}
It follows that equation (\ref{eq:rt_hform}) becomes
\begin{equation}
\frac{\partial K_-}{\partial \tau} = \bar{\mu} F_+ \left( 1 - \omega_0 g_0 \right).
\label{eq:rt_kform}
\end{equation}
Note that all of these steps used to derive ${\cal G}$ and ${\cal I}$, for ${\cal H}=\mu$, are invalid if the integration is not carried out over all $4\pi$ steradians.  Partial integration is akin to making specific assumptions about the asymmetry properties of ${\cal P}$.

\subsubsection{Consistency with Isotropic Expressions}

By adopting the appropriate closures, equations (\ref{eq:rt_hform2}) and (\ref{eq:rt_kform}) may be transformed into a pair of equations for the outgoing and incoming fluxes.  A basic consistency check is to demand that they reduce to the pair of equations in (\ref{eq:twostream_iso_2}) when $g_0=0$ (isotropic scattering).

Invoking the first Eddington coefficient, $\epsilon \equiv F_+/J$, we see that equation (\ref{eq:rt_hform2}) reduces to the second equation in (\ref{eq:twostream_iso_2}) if $\epsilon = 1/2$.  To recover the first equation in (\ref{eq:twostream_iso_2}), we need to define a second Eddington coefficient,
\begin{equation}
\epsilon_2 \equiv \frac{K_-}{F_+},
\end{equation}
which is assumed to be constant.  In the $g_0=0$ limit, equation (\ref{eq:rt_kform}) reduces to the first equation in (\ref{eq:twostream_iso_2}) only if 
\begin{equation}
\epsilon_2 = \frac{F_+}{2 F_-}.
\label{eq:eddington2}
\end{equation}
The fact that the second Eddington coefficient may be stated in terms of the computed quantities ($F_\uparrow$ and $F_\downarrow$) implies that one may use it to check if the vertical resolution of one's model atmosphere is sufficient.  Analogous to the isothermal assumption on $B$, we are asserting that $\epsilon_2$ is constant within each atmospheric layer.

Note that our method of derivation differs from the textbook treatments of \cite{gy89} and \cite{pierrehumbert}.  In \cite{gy89}, it is assumed that two characteristic, constant values of the intensity may be defined: $I_\uparrow$ and $I_\downarrow$.  It is then assumed that $F_- = \pi (I_\uparrow - I_\downarrow)$ and $J=2\pi(I_\uparrow + I_\downarrow)$; see their equations (2.142) and (2.143).  The two-stream equations are then derived assuming $F_\uparrow = \pi I_\uparrow$ and $F_\downarrow = \pi I_\downarrow$.  In \cite{pierrehumbert}, the method of using $\epsilon_2$ is mentioned, but never explicitly executed; the correspondence to the isotropic limit, in the manner we have presented it, is not discussed.  Furthermore, the expression for the second Eddington coefficient, in equation (\ref{eq:eddington2}), is not derived.

\subsubsection{Governing Equations in Two-Stream Form}

With this pair of closures, the total and net fluxes are governed by the equations,
\begin{equation}
\begin{split}
\frac{\partial F_+}{\partial \tau} =& \left( \gamma_{\rm a} + \gamma_{\rm s} \right) F_-, \\
\frac{\partial F_-}{\partial \tau} =& \left( \gamma_{\rm a} - \gamma_{\rm s} \right) F_+ - 2\gamma_{\rm B} B,
\end{split}
\label{eq:twostream_non_iso_1}
\end{equation}
where we have defined
\begin{equation}
\begin{split}
\gamma_{\rm a} &= \bar{\mu} \left[ 2 - \omega_0 \left( 1 + g_0 \right) \right], \\
\gamma_{\rm s} &= \bar{\mu} \omega_0 \left( 1 - g_0 \right), \\
\gamma_{\rm B} &= 2 \pi \bar{\mu} \left( 1 - \omega_0 \right). \\
\end{split}
\label{eq:twostream_non_isotropic_coefficients}
\end{equation}
The incoming and outgoing fluxes obey the same mathematical form as given in equation (\ref{eq:twostream_iso}), but with the coefficients given in equation (\ref{eq:twostream_non_isotropic_coefficients}).  By ensuring correspondence with the isotropic limit, these coefficients obey energy conservation in the purely scattering limit and reproduce the correct blackbody flux in the purely absorbing limit for an opaque atmosphere.  It is worth noting that the constraint of energy conservation in the purely scattering limit is independent of $g_0$.

The set of coefficients in equation (\ref{eq:twostream_non_isotropic_coefficients}) is traditionally known as the ``hemispheric" or ``hemi-isotropic" closure \citep{mw80,toon89,pierrehumbert}. While its statement is certainly not novel, our derivation of the hemispheric closure is firmly grounded by a desire to ensure energy conservation.  We find that the hemispheric closure derives naturally from the radiative transfer equation.

\subsubsection{General Solution with Non-Isotropic Scattering and Non-Isothermal Layers}
\label{subsect:non_iso_non_isothermal}

Manipulating the pair of equations in (\ref{eq:twostream_non_iso_1}) yields
\begin{equation}
\frac{\partial^2 F_+}{\partial \tau^2} - \alpha^2 F_+ = - 2 \gamma_{\rm B} \left( \gamma_{\rm a} + \gamma_{\rm s} \right) B,
\label{eq:twostream_non_iso_2}
\end{equation}
but with a more general definition for $\alpha$,
\begin{equation}
\alpha \equiv \left[ \left( \gamma_{\rm a} + \gamma_{\rm s} \right) \left( \gamma_{\rm a} - \gamma_{\rm s} \right) \right]^{1/2}.
\label{eq:alpha_2}
\end{equation}

The method for solving equation (\ref{eq:twostream_non_iso_2}) has previously been described in \S\ref{subsect:general_solution_iso}.  Here, we simply state our results for non-isothermal layers with the Planck function as given by equation (\ref{eq:b_expand}).  The total and net fluxes are
\begin{equation}
\begin{split}
F_+ =& {\cal A}_1 \exp{\left( \alpha \tau \right)} + {\cal A}_2 \exp{\left( -\alpha \tau \right)} + \frac{2 \gamma_{\rm B} B}{\gamma_{\rm a} - \gamma_{\rm s}}, \\
F_- =& \left( \frac{\gamma_{\rm a} - \gamma_{\rm s}}{\gamma_{\rm a} + \gamma_{\rm s}} \right)^{1/2} \left[ {\cal A}_1 \exp{\left( \alpha \tau \right)} - {\cal A}_2 \exp{\left( -\alpha \tau \right)} \right] \\
&+ \frac{2 \gamma_{\rm B} \bar{\mu} B^\prime}{\left( \gamma_{\rm a} + \gamma_{\rm s} \right) \left( \gamma_{\rm a} - \gamma_{\rm s} \right)},
\end{split}
\end{equation}
from which the outgoing and incoming fluxes may be obtained,
\begin{equation}
\begin{split}
F_{\uparrow} =& {\cal A}_1 \zeta_+ \exp{\left( \alpha \tau \right)} + {\cal A}_2 \zeta_- \exp{\left( -\alpha \tau \right)} \\
&+ \frac{\gamma_{\rm B}}{\gamma_{\rm a} - \gamma_{\rm s}} \left( B + \frac{\bar{\mu} B^\prime}{\gamma_{\rm a} + \gamma_{\rm s}} \right), \\
F_{\downarrow} =& {\cal A}_1 \zeta_- \exp{\left( \alpha \tau \right)} + {\cal A}_2 \zeta_+ \exp{\left( -\alpha \tau \right)} \\
&+ \frac{\gamma_{\rm B}}{\gamma_{\rm a} - \gamma_{\rm s}} \left( B - \frac{\bar{\mu} B^\prime}{\gamma_{\rm a} + \gamma_{\rm s}} \right), \\
\end{split}
\end{equation}
which have a more general definition of the $\zeta_+$ and $\zeta_-$ coefficients,
\begin{equation}
\begin{split}
\zeta_\pm &\equiv \frac{1}{2} \left[ 1 \pm \left( \frac{\gamma_{\rm a} - \gamma_{\rm s}}{\gamma_{\rm a} + \gamma_{\rm s}} \right)^{1/2} \right] \\
&= \frac{1}{2} \left[ 1 \pm \left( \frac{1-\omega_0}{1-\omega_0 g_0} \right)^{1/2} \right].
\end{split}
\end{equation}
It is reassuring that $\zeta_\pm$ does not diverge as $\omega_0, g_0 \rightarrow 0$ or $\omega_0, g_0 \rightarrow 1$.  The quadrature closure gives the same expression for $\zeta_\pm$ (see Table 3 for exact forms of $\gamma_{\rm a}$ and $\gamma_{\rm s}$).  The Eddington closure adds a factor of 2/3 within the square root.  All of the closures do not display divergence.

By again imposing the boundary conditions $F_{\uparrow_2}$ and $F_{\downarrow_1}$, we may derive the outgoing and incoming fluxes from a pair of atmospheric layers in terms of the transmission function, 
\begin{equation}
\begin{split}
F_{\uparrow_1} =& \frac{1}{\left( \zeta_- {\cal T} \right)^2 - \zeta_+^2} \left\{ \left( \zeta_-^2 - \zeta_+^2 \right) {\cal T} F_{\uparrow_2} - \zeta_- \zeta_+ \left( 1 - {\cal T}^2 \right) F_{\downarrow_1} \right. \\
&+ \left. \frac{\gamma_{\rm B}}{\gamma_{\rm a} - \gamma_{\rm s}} \left[ {\cal B}_{1+} \left( \zeta^2_- {\cal T}^2 - \zeta_+^2 \right) + {\cal B}_{2+} {\cal T}  \left( \zeta_+^2 - \zeta_-^2 \right) \right. \right. \\
&+ \left. \left. {\cal B}_{1-} \zeta_- \zeta_+ \left( 1 - {\cal T}^2 \right) \right] \right\},  \\
F_{\downarrow_2} =& \frac{1}{\left( \zeta_- {\cal T} \right)^2 - \zeta_+^2} \left\{ \left( \zeta_-^2 - \zeta_+^2 \right) {\cal T} F_{\downarrow_1} - \zeta_- \zeta_+ \left( 1 - {\cal T}^2 \right) F_{\uparrow_2} \right. \\
&+ \left. \frac{\gamma_{\rm B}}{\gamma_{\rm a} - \gamma_{\rm s}} \left[ {\cal B}_{2-} \left( \zeta^2_- {\cal T}^2 - \zeta_+^2 \right) + {\cal B}_{1-} {\cal T}  \left( \zeta_+^2 - \zeta_-^2 \right) \right. \right. \\
&+ \left. \left. {\cal B}_{2+} \zeta_- \zeta_+ \left( 1 - {\cal T}^2 \right) \right] \right\},  \\
\end{split}
\label{eq:twostream_iso_general_solution_non_isothermal_2}
\end{equation}
where the expression for ${\cal B}_{i\pm}$ is now generalized to
\begin{equation}
\begin{split}
{\cal B}_{i-} &\equiv B_1 + B^\prime \bar{\mu} \left( \tau_i - \tau_1 \right) - \frac{\bar{\mu} B^\prime}{\gamma_{\rm a} + \gamma_{\rm s}}, \\
{\cal B}_{i+} &\equiv B_2 + B^\prime \bar{\mu} \left( \tau_i - \tau_2 \right) + \frac{\bar{\mu} B^\prime}{\gamma_{\rm a} + \gamma_{\rm s}}.
\end{split}
\end{equation}

In the limit of isothermal layers (${\cal B}_{i\pm}=B=B_0$), we obtain
\begin{equation}
\begin{split}
F_{\uparrow_1} =& \frac{1}{\left( \zeta_- {\cal T} \right)^2 - \zeta_+^2} \left\{ \left( \zeta_-^2 - \zeta_+^2 \right) {\cal T} F_{\uparrow_2} - \zeta_- \zeta_+ \left( 1 - {\cal T}^2 \right) F_{\downarrow_1} \right. \\
&+ \left. \frac{\gamma_{\rm B} B}{\gamma_{\rm a} - \gamma_{\rm s}} \left[ \zeta_- \zeta_+ \left( 1 - {\cal T}^2 \right) - \left( \zeta_-^2 {\cal T} + \zeta_+^2 \right) \left( 1 - {\cal T} \right) \right] \right\},  \\
F_{\downarrow_2} =& \frac{1}{\left( \zeta_- {\cal T} \right)^2 - \zeta_+^2} \left\{ \left( \zeta_-^2 - \zeta_+^2 \right) {\cal T} F_{\downarrow_1} - \zeta_- \zeta_+ \left( 1 - {\cal T}^2 \right) F_{\uparrow_2} \right. \\
&+ \left. \frac{\gamma_{\rm B} B}{\gamma_{\rm a} - \gamma_{\rm s}} \left[ \zeta_- \zeta_+ \left( 1 - {\cal T}^2 \right) - \left( \zeta_-^2 {\cal T} + \zeta_+^2 \right) \left( 1 - {\cal T} \right) \right] \right\}.  \\
\end{split}
\label{eq:twostream_iso_general_solution_non_isothermal}
\end{equation}

In the limit of pure scattering ($\omega_0=1$), equations (\ref{eq:twostream_iso_general_solution_non_isothermal_2}) and (\ref{eq:twostream_iso_general_solution_non_isothermal}) need to be replaced by
\begin{equation}
\begin{split}
F_{\uparrow_1} &= F_{\uparrow_2} - \frac{\left( F_{\uparrow_2} - F_{\downarrow_1} \right) \left( \gamma_{\rm a} + \gamma_{\rm s} \right) \left( \tau_2 - \tau_1 \right)}{2 + \left( \gamma_{\rm a} + \gamma_{\rm s} \right) \left( \tau_2 - \tau_1 \right)}, \\
F_{\downarrow_2} &= F_{\downarrow_1} + \frac{\left( F_{\uparrow_2} - F_{\downarrow_1} \right) \left( \gamma_{\rm a} + \gamma_{\rm s} \right) \left( \tau_2 - \tau_1 \right)}{2 + \left( \gamma_{\rm a} + \gamma_{\rm s} \right) \left( \tau_2 - \tau_1 \right)}. \\
\end{split}
\end{equation}

Our statement of these solutions in terms of the coefficients $\gamma_{\rm a}$, $\gamma_{\rm s}$ and $\gamma_{\rm B}$ allows for other closures to be considered (see \S\ref{subsect:other_closures}).

\subsection{Transitioning from Two-Stream Treatment to Flux-Limited Diffusion}
\label{subsect:fld}

If one specifies a sufficient number of model layers, the two-stream treatment is a good approximation at optical depths of order unity or less.  When the optical depth becomes large, a prohibitive number of layers may be needed.  Deep within an exoplanet, the passage of radiation resembles diffusion and the total and net fluxes depend on having $\partial B/\partial \tau \ne 0$ \citep{mihalas}.  One needs a way to transition from the two-stream treatment to the diffusion approximation.  Physically, the transition occurs where the photon mean free path becomes much smaller than the vertical spatial resolution.

A fundamental problem with approximating radiative transfer by diffusion is that the diffusion equation does not obey causality, i.e., it will formally allow superluminal motion.  It has been remedied by the invention of ``flux-limited diffusion", where transport is limited by the speed of light \citep{lp81,narayan92}.  Flux-limited diffusion produces the correct behavior in the optically thin and thick limits, but its accuracy when $\tau_0 \sim 1$ is suspect \citep{mm99}.  Its use may be abandoned altogether by considering the fully time-dependent radiative transfer equation \citep{mm99}.  Since we are employing diffusion only when $\tau_0 \gg 1$, our approach is ``flux-limited" by definition, while benefitting from the accuracy of the two-stream approximation at $\tau_0 \lesssim 1$.

Our starting points are equations (\ref{eq:rt_hform2}) and (\ref{eq:rt_kform}).  To close this pair of equations, we define the third Eddington coefficient,
\begin{equation}
\epsilon_3 \equiv \frac{K_-}{J}.
\end{equation}
In the deep interior, we assert that the intensity field becomes Planckian, scattering becomes isotropic and total and net quantities become equal,\footnote{If we insist on the two-stream interpretation, then it means the total and outgoing fluxes are also equal.}
\begin{equation}
J = 4\pi B, ~g_0 = 0, ~F_+ = F_- = \pi B_{\rm int}, ~K_+ = K_-,
\end{equation}
where $B_{\rm int} \equiv B(T_{\rm int})$.  The interior heat of the exoplanet is represented by an internal temperature, $T_{\rm int}$.  It follows that $\epsilon_3 = K_+ / 4\pi B = 1/3$ if $I=B$.

In this limit, we obtain
\begin{equation}
\begin{split}
&F_\pm = \frac{\partial K_\pm}{\partial \tau_0} = 4\pi \epsilon_3 \frac{\partial B}{\partial \tau_0}, \\
&\frac{\partial F_\pm}{\partial \tau_0} = 0.
\end{split}
\end{equation}
By definition, heating in the deep interior is in radiative equilibrium ($\partial F_- / \partial \tau_0 = 0$).  It is apparent that if the isothermal approximation is made ($\partial B / \partial \tau_0 = 0$), then interior heating is missed altogether ($F_- = 0$).  We define the wavelength-integrated quantities,
\begin{equation}
\begin{split}
{\cal K}_\pm &\equiv \int^\infty_0 K_\pm ~d\lambda, \\
{\cal F}_\pm &\equiv \int^\infty_0 F_\pm ~d\lambda, \\
\end{split}
\end{equation}
and assert that a wavelength-integrated, average opacity exists such that the following expression is true \citep{mihalas,mm99},
\begin{equation}
\int^\infty_0 \frac{1}{\kappa} \frac{\partial K_\pm}{\partial z} ~d\lambda = {\cal F}_\pm = \frac{1}{\kappa_{\rm R}} \frac{\partial {\cal K}_\pm}{\partial z},
\end{equation}
where $d \tau_0 \equiv \rho\kappa dz$, $\kappa$ is the wavelength-dependent total\footnote{Includes both absorption and scattering, otherwise known as the ``extinction opacity".} opacity, $\rho$ is the mass density and $z$ is the vertical spatial coordinate, from which the definition of the Rosseland mean opacity follows,
\begin{equation}
\kappa_{\rm R} \equiv \frac{4 \sigma_{\rm SB} T^3}{\pi} \left(\int \frac{1}{\kappa} \frac{\partial B}{\partial T} ~d\lambda\right)^{-1},
\label{eq:rosseland}
\end{equation}
with the gradient of the Planck function being
\begin{equation}
\frac{\partial B}{\partial T} = \frac{B^2 \lambda^4}{2 c k_{\rm B} T^2} ~\exp{\left( \frac{hc}{\lambda k_{\rm B} T} \right)},
\end{equation}
where $\lambda$ is the wavelength, $k_{\rm B}$ is Boltzmann's constant, $T$ is the temperature and $h$ is the Planck constant.  The $\partial/\partial T$ operation in equation (\ref{eq:rosseland}) cannot be taken out of the integral because $\kappa$ generally depends on temperature.  Note that the definition for $\kappa_{\rm R}$ does not depend on $\epsilon_3$.

The total heat content (${\cal F}_+$) and net heating (${\cal F}_-$) of the deep interior is
\begin{equation}
{\cal F}_\pm = \frac{16 \epsilon_3 g \sigma_{\rm SB} T^3 }{\kappa_{\rm R}} \frac{\partial T}{\partial P} = \sigma_{\rm SB} T^4_{\rm int},
\label{eq:fick}
\end{equation}
where $g$ is the surface gravity of the exoplanet and hydrostatic equilibrium has been assumed.  Equation (\ref{eq:fick}) takes the same mathematical form as Fick's law of diffusion, where the flux is proportional to a diffusion coefficient and the gradient of an internal quantity.  By integrating equation (\ref{eq:fick}), one obtains
\begin{equation}
T = \left[ \frac{1}{4 \epsilon_3} \left( \tau_{\rm R} + {\cal C} \right) \right]^{1/4} T_{\rm int},
\label{eq:tdeep}
\end{equation}
where the Rosseland mean optical depth is
\begin{equation}
\tau_{\rm R} \equiv \frac{1}{g} \int \kappa_{\rm R} dP,
\end{equation}
and ${\cal C}$ is a constant of integration.  Equation (\ref{eq:tdeep}) is exactly Milne's solution for self-luminous atmospheres \citep{mihalas,mm99}, where the internal temperature is boosted by a factor, involving the optical depth, at large pressures.  In \S\ref{sect:tp}, we will see that ${\cal C} = 8/9$, when we examine analytical solutions of the temperature-pressure profile.

A few potential concerns are worth elucidating.  When applied sharply to specific wavelengths, the validity of the diffusion approximation is suspect, since we expect absorption and re-emission to be non-coherent.  However, the diffusion approximation is reasonable when it is applied to a collection of wavelength bins, where the width of each bin is much larger than the typical width of a spectral line.  The transition to the diffusion approximation occurs in a wavelength-independent manner as determined by the onset of the deep temperature-pressure profile, as stated in equation (\ref{eq:tdeep}), but it is worth noting that the Rosseland mean opacity is weighted towards \textit{lower} opacities \citep{mihalas}.  Physically, this means that at wavelengths where the atmosphere is the most transparent, $\tau_0 \sim 1$ and $\tau_{\rm R} \sim 1$ occur essentially at the same depth or pressure.  

\section{Temperature-Pressure Profiles with Non-Isotropic, Coherent Scattering}
\label{sect:tp}

We generalize the work of \cite{guillot10} (pure absorption) and \cite{hhps12} (isotropic scattering) by including non-isotropic scattering and a non-constant shortwave opacity in our derivation of the analytical temperature-pressure profiles.  Additionally, we distinguish between total and net fluxes and resolve several lingering issues in \cite{guillot10} and \cite{hhps12}.

We adopt the dual-band approximation, where incident stellar irradiation and thermal emission from the exoplanetary atmosphere reside in the ``shortwave" (denoted by ``S") and ``longwave" (denoted by ``L"), respectively.  We define several quantities that are integrated over the shortwave and longwave,
\begin{equation}
\begin{split}
&J_{\rm S} \equiv \int_{\rm S} J ~d\lambda, ~F_{\rm S} \equiv \int_{\rm S} F_- ~d\lambda, ~K_{\rm S} \equiv \int_{\rm S} K_- ~d\lambda, \\
&J_{\rm L} \equiv \int_{\rm L} J ~d\lambda, ~F_{\rm L} \equiv \int_{\rm L} F_- ~d\lambda, ~K_{\rm L} \equiv \int_{\rm L} K_- ~d\lambda. \\
\end{split}
\end{equation}

In this section, we require two additional Eddington coefficients,
\begin{equation}
\epsilon_{\rm L} \equiv \frac{F_{\rm L}}{J_{\rm L}}, ~\epsilon_{\rm L_3} \equiv \frac{K_{\rm L}}{J_{\rm L}}.
\end{equation}
By requiring that it corresponds to $\epsilon_3$, we set $\epsilon_{\rm L_3}=1/3$.  Using our existing definitions and values for $\epsilon$, $\epsilon_2$ and $\epsilon_3$ (see Table 2), we have
\begin{equation}
\epsilon_{\rm L} = \frac{\epsilon^2}{2 \epsilon_3} = \frac{3}{8}.
\end{equation}
Note that \cite{guillot10} and \cite{hhps12} set $\epsilon_{\rm L}=1/2$.

\subsection{General Equations and Energy Conservation}

We begin with equations (\ref{eq:rt_hform2}) and (\ref{eq:rt_kform}), the intermediate form of the governing equations with non-isotropic scattering that leads to the two-stream and flux-limited-diffusion treatments.  Instead of using the optical depth as the independent variable, we write
\begin{equation}
d\tau_0 = \kappa ~dm = \frac{\kappa_{\rm a}}{1 - \omega_0} ~dm,
\end{equation}
with $m$ being the column mass.  In hydrostatic equilibrium, we have $P = mg$.  Formulating the equations in terms of $m$, a wavelength-independent quantity, will later allow us to define separate shortwave and longwave opacities.  Instead of using the total/extinction opacity ($\kappa$), we have used the absorption opacity ($\kappa_{\rm a}$) as this allows us to cleanly separate out the component due to scattering in the form of the single-scattering albedo ($\omega_0$).  Equations (\ref{eq:rt_hform2}) and (\ref{eq:rt_kform}) become
\begin{equation}
\begin{split}
\frac{\partial F_-}{\partial m} &= \kappa_{\rm a} \left( J - 4 \pi B \right), \\
\frac{\partial K_-}{\partial m} &= \frac{ \kappa_{\rm a} F_+}{\beta_0^2}, \\
\end{split}
\label{eq:tp_govern}
\end{equation}
where we have defined
\begin{equation}
\beta_0 \equiv \left( \frac{1 - \omega_0}{1 - \omega_0 g_0} \right)^{1/2}.
\end{equation}
Previously, \cite{guillot10} and \cite{hhps12} wrote down less general forms of equation (\ref{eq:tp_govern}) with total, instead of net, quantities.

The first equation in (\ref{eq:tp_govern}) allows the conservation of energy to be expressed,
\begin{equation}
\int^\infty_0 \frac{\partial F_-}{\partial m} d\lambda = \frac{\partial {\cal F}_-}{\partial m} = Q = \kappa_{\rm S} J_{\rm S} + \kappa_{\rm L} \left( J_{\rm L} - 4 \sigma_{\rm SB} T^4 \right).
\label{eq:energy_conserve}
\end{equation}
We will properly define the shortwave and longwave opacities, denoted respectively by $\kappa_{\rm S}$ and $\kappa_{\rm L}$, shortly.  The heating rate is given by $Q$.  Radiative equilibrium is obtained when $\partial {\cal F}_0 / \partial m = Q = 0$.  Note that this interpretation differs from that of \cite{guillot10} and \cite{hhps12}, who interpreted quantities associated with $Q$, integrated over all angles, to vanish because of conservative heat transport.

By integrating equation (\ref{eq:energy_conserve}) over column mass, we obtain
\begin{equation}
\int^\infty_m \frac{\partial {\cal F}_-}{\partial m} ~dm = \tilde{Q}\left(m,\infty\right),
\end{equation}
where
\begin{equation}
\tilde{Q}\left(m_1,m_2\right) \equiv \int^{m_2}_{m_1} Q ~dm,
\end{equation}
from which it follows that
\begin{equation}
\begin{split}
F_{\rm L} &= {\cal F}_\infty - F_{\rm S} - \tilde{Q}\left(m,\infty\right), \\
J_{\rm L_0} &= \frac{1}{\epsilon_{\rm L}} \left[ {\cal F}_\infty - F_{\rm S_0} - \tilde{Q}\left(0,\infty\right) \right].
\end{split}
\label{eq:energy_conserve_2}
\end{equation}
We have defined $F_{\rm S_0} \equiv F_{\rm S}(m=0)$, $F_{\rm L_0} \equiv F_{\rm L}(m=0)$ and $J_{\rm L_0} \equiv J_{\rm L}(m=0)$.  The quantity ${\cal F}_\infty$ is the bolometric net flux from the deep interior (as $m \rightarrow \infty$),
\begin{equation}
{\cal F}_\infty = \sigma_{\rm SB} T^4_{\rm int}.
\end{equation}

\subsection{Shortwave}

The shortwave refers to the range of wavelengths where incident starlight is the dominant source of energy.  It usually occurs in the optical.

\subsubsection{Shortwave Closure, the Collimated Beam Approximation and the Bond Albedo}
\label{subsect:shortwave_albedo}

Before we derive the shortwave equations and their solutions, we need to relate $J_{\rm S}$ and $K_{\rm S}$ via a shortwave closure relation.  Previously, \cite{guillot10} assumed that
\begin{equation}
\epsilon_{\rm S} \equiv \frac{K_{\rm S}}{J_{\rm S}} = \mu^2.
\label{eq:shortwave_closure}
\end{equation}
\cite{hhps12} tried to justify this closure via the collimated beam approximation,
\begin{equation}
I_{\rm S} = I_{\uparrow S} \delta\left(\mu^\prime - \mu\right) + I_{\downarrow S} \delta\left(\mu^\prime + \mu\right).
\label{eq:coll_approx}
\end{equation}
When one does not distinguish between total and net quantities, one can simultaneously satisfy equation (\ref{eq:shortwave_closure}) and the identity in (\ref{eq:shortwave_identity}); we will derive the latter later.  In our current, improved formulation, this is no longer possible.  

Such a finding has several implications.  First, it means that equation (\ref{eq:shortwave_closure}) will have to be justified after the fact, upon obtaining the solution for $F_{\rm S}$.  We will see that this closure correctly produces Beer's law.  

Second, it implies that the expression for the Bond albedo previously derived by \cite{hhps12} using the collimated beam approximation, $A_{\rm B} = (1 - \sqrt{1-\omega_0}) / (1 + \sqrt{1+\omega_0})$, may no longer be self-consistent within our improved formalism.  However, we may directly derive the spherical albedo ($A_{\rm s}$) from our two-stream solutions with non-isotropic scattering, previously stated in equation (\ref{eq:twostream_iso_general_solution_non_isothermal}).  We may then integrate $A_{\rm s}$ over the shortwave to obtain the Bond albedo.  If we set $F_{\uparrow_2}=0$ and $B=0$ and integrate over the shortwave, we obtain
\begin{equation}
\begin{split}
A_{\rm s} &\equiv \frac{F_{\uparrow_1}}{F_{\downarrow_1}} = \frac{\zeta_-}{\zeta_+} = \frac{1 - \beta_0}{1 + \beta_0}, \\
A_{\rm B} &\equiv \int_{\rm S}  A_{\rm s} ~d\lambda = \frac{1 - \beta_{\rm S_0}}{1 + \beta_{\rm S_0}},
\end{split}
\label{eq:bond_albedo}
\end{equation}
where $\beta_{\rm S_0}$ is the value of $\beta_0$ in the shortwave, which we will describe more carefully in equation (\ref{eq:beta_s0}).  Note that equation (\ref{eq:bond_albedo}) was derived for an opaque atmosphere (${\cal T}=0$).  Physically, one is asserting that when scattering is absent, all of the incident stellar irradiation is completely absorbed.  Coincidentally, equation (\ref{eq:bond_albedo}) is identical to the expression derived by \cite{hhps12} in the limit of isotropic scattering.  The functional behaviors of $A_{\rm s}$ and $A_{\rm B}$ are shown in Figure \ref{fig:zetas} via the curves of $\zeta_-/\zeta_+$.

We find it useful to express the quantity $\beta_{\rm S_0}$ in terms of the Bond albedo,
\begin{equation}
\beta_{\rm S_0} = \frac{1 - A_{\rm B}}{1 + A_{\rm B}}.
\label{eq:bond_albedo_2}
\end{equation}
Degenerate combinations of the single-scattering albedo and asymmetry factor may produce the same Bond albedo.

\cite{hd13} have previously derived an expression for $A_{\rm B}$ involving non-isotropic, coherent scattering, by generalizing the approach of \cite{pierrehumbert}.  In these approaches, an additional ``direct beam" term was added to the source term ($S$) in the radiative transfer equation to account for heating by incident starlight \citep{chandra60}.  Given that the two-stream approximation is a one-dimensional treatment, we feel that regarding the solutions in equation (\ref{eq:twostream_iso_general_solution_non_isothermal}) as being wavelength-dependent and using the $F_{\downarrow_1}$ boundary condition to account for stellar irradiation, across wavelength, is sufficient and that a direct beam term is superfluous \citep{mw80}.

\subsubsection{Shortwave Equations and Solutions}

Integrating the equations in (\ref{eq:tp_govern}) over the shortwave, we obtain
\begin{equation}
\begin{split}
\frac{\partial F_{\rm S}}{\partial m} &= \kappa_{\rm S} J_{\rm S}, \\
\frac{\partial K_{\rm S}}{\partial m} &= \frac{ \kappa_{\rm S}^\prime F_{\rm S}}{\beta_{\rm S_0}^2}, \\
\end{split}
\label{eq:sw_eqn}
\end{equation}
where the absorption mean opacity is
\begin{equation}
\kappa_{\rm S} \equiv \frac{\int_{\rm S} \kappa_{\rm a} J ~d\lambda}{\int_{\rm S} J ~d\lambda}.
\label{eq:abs_mean}
\end{equation}
In a departure from its traditional definition, the flux mean opacity is
\begin{equation}
\kappa_{\rm S}^\prime \equiv \frac{\int_{\rm S} \kappa_{\rm a} F_- ~d\lambda}{\int_{\rm S} F_- ~d\lambda}.
\label{eq:flux_mean}
\end{equation}
Usually, the flux mean opacity is defined using $\kappa$ instead of $\kappa_{\rm a}$ \citep{mm99}.  Our approach comes about because we have approximated 
\begin{equation}
\beta_{\rm S_0} = \left( \frac{1 - \omega_{\rm S_0}}{1 - \omega_{\rm S_0} g_{\rm S_0}} \right)^{1/2}
\label{eq:beta_s0}
\end{equation}
as being constant with wavelength, such that $\omega_{\rm S_0}$ and $g_{\rm S_0}$ are constant, representative values of the single-scattering albedo and asymmetry factor, respectively, in the shortwave.

In order to combine the equations in (\ref{eq:sw_eqn}), we assume that $\kappa_{\rm S} = \kappa_{\rm S}^\prime$.  We shall simply call $\kappa_{\rm S}$ the ``shortwave opacity".  It follows that
\begin{equation}
\begin{split}
&\frac{\partial^2 J_{\rm S}}{\partial m^2} - \frac{1}{\kappa_{\rm S}} \frac{\partial \kappa_{\rm S}}{\partial m} \frac{\partial J_{\rm S}}{\partial m} - \left( \frac{\kappa_{\rm S}}{\mu \beta_{\rm S_0}} \right)^2 J_{\rm S} = 0,\\
&\frac{\partial^2 F_{\rm S}}{\partial m^2} - \frac{1}{\kappa_{\rm S}} \frac{\partial \kappa_{\rm S}}{\partial m} \frac{\partial F_{\rm S}}{\partial m} - \left( \frac{\kappa_{\rm S}}{\mu \beta_{\rm S_0}} \right)^2 F_{\rm S} = 0.
\end{split}
\label{eq:tp_2ode}
\end{equation}

If we assume the shortwave opacity to take the form,
\begin{equation}
\kappa_{\rm S} = \kappa_{\rm S_0} \left( \frac{m}{m_0} \right)^n,
\label{eq:kappa_s}
\end{equation}
where $\kappa_{\rm S_0}$ is its value at the bottom of the model domain, $n$ is a dimensionless index, $m_0 = P_0 / g$ and $P_0$ is the pressure at the bottom of the model domain, then we obtain
\begin{equation}
\begin{split}
J_{\rm S} &= J_{\rm S_0} \exp{\left( \frac{\beta_{\rm S}}{\mu} \right)}, \\
F_{\rm S} &= F_{\rm S_0} \exp{\left( \frac{\beta_{\rm S}}{\mu} \right)}, \\
\end{split}
\label{eq:shortwave_solutions}
\end{equation}
with $J_{\rm S_0} \equiv J_{\rm S}(m=0)$ and
\begin{equation}
\beta_{\rm S} \equiv \frac{\kappa_{\rm S} m}{\left( n + 1 \right) \beta_{\rm S_0}}.
\label{eq:beta_s}
\end{equation}
The expressions in (\ref{eq:shortwave_solutions}) generalize Beer's law.  We have picked the solution branch with the positive exponent, because we have $-1 \le \mu \le 0$ and we require that $J_{\rm S}, F_{\rm S} \rightarrow 0$ as $m \rightarrow \infty$.  It follows that
\begin{equation}
F_{\rm S} = \mu \beta_{\rm S_0} J_{\rm S}.
\label{eq:shortwave_identity}
\end{equation}

\subsubsection{Photon Deposition Depth}

The shortwave flux at $m=0$ is interpreted as the incident stellar flux,
\begin{equation}
F_{\rm S_0} = \mu F_\star,
\end{equation}
where the ``stellar constant"\footnote{Generalized from the ``solar constant".} is
\begin{equation}
F_\star \equiv 
\begin{cases}
\sigma_{\rm SB} T^4_{\rm irr}, & 0 \le \phi \le \pi, \\
0, & \pi \le \phi \le 2\pi,
\end{cases}
\end{equation}
and the irradiation temperature is
\begin{equation}
T_{\rm irr} = T_\star \left( \frac{R_\star}{a} \right)^{1/2} \left( 1 - A_{\rm B} \right)^{1/4},
\end{equation}
with $T_\star$ being the effective stellar temperature, $R_\star$ the stellar radius and $a$ the distance between the star and the exoplanet.  It is important to note that $F_{\rm S_0} < 0$ arises naturally from the fact that it is a \textit{net flux with a vanishing outgoing component.}  No arbitrary adjustments of signs are necessary, as was done in \cite{guillot10} and \cite{hhps12}.

By using the expression for $F_{\rm S}$ from equation (\ref{eq:shortwave_solutions}), we find that
\begin{equation}
\bar{F}_{\rm S} \equiv \frac{1}{2\pi} \int^{2\pi}_0 \int^0_{-1} F_{\rm S} ~d\mu ~d\phi = - \frac{\sigma_{\rm SB} T^4_{\rm irr} {\cal E}_3}{2},
\end{equation}
where ${\cal E}_3 = {\cal E}_3(\beta_{\rm S})$ and the exponential integral of the $i$-th order is defined as \citep{abram,aw95}
\begin{equation}
{\cal E}_i \left( y \right) \equiv \int^\infty_1 x^{-i} \exp{\left(-xy \right)} ~dx.
\label{eq:expint}
\end{equation}
It follows that 
\begin{equation}
\frac{\bar{F}_{\rm S}}{\bar{F}_{\rm S_0}} = 2 {\cal E}_3.
\end{equation}

The photon deposition depth is defined as the pressure level where $\bar{F}_{\rm S}/\bar{F}_{\rm S_0}$ suffers one e-folding, i.e., is equal to about 0.368 \citep{hhps12}.  Physically, this is the pressure level at which most of the incident starlight is being absorbed ($P_{\rm D}$).  This occurs when $\beta_{\rm S} \approx 0.63$.  It follows that
\begin{equation}
\begin{split}
P_{\rm D} &= \left[ \frac{0.63 \left( n + 1 \right) g P_0^n}{\kappa_{\rm S_0}} \right]^{1/\left(n+1\right)} \left( \frac{1 - \omega_{\rm S_0}}{1 - \omega_{\rm S_0} g_{\rm S_0}} \right)^{1/2\left(n+1\right)} \\
&= \left[ \frac{0.63 \left( n + 1 \right) g P_0^n}{\kappa_{\rm S_0}} \right]^{1/\left(n+1\right)} \left( \frac{1 - A_{\rm B}}{1 + A_{\rm B}} \right)^{1/\left(n+1\right)}.
\end{split}
\label{eq:pdd}
\end{equation}
It has the expected physical property that, as the scattering becomes more backward-peaked ($g_{\rm S_0} < 0$), the photon deposition depth resides higher in the atmosphere.  As $n \rightarrow \infty$, $P_{\rm D} \rightarrow P_0$.  Equation (\ref{eq:pdd}) generalizes the expression derived by \cite{hhps12} in the limit of isotropic, coherent scattering and $n=0$.  

When $n=0$, the expression for $P_{\rm D}$ is particularly useful because it is independent of $P_0$.  Figure \ref{fig:pdd} shows calculations of $P_{\rm D}$ (with $n=0$) as a function of $\omega_0$ for different values of $g_0$.  For pure forward scattering ($g_{\rm S_0}=1$), photon deposition behaves as if one is in the purely absorbing limit.  Backward scattering ($g_{\rm S_0}=-1$) tends to raise the photon deposition depth to higher altitudes (lower pressures).  

\begin{figure}
\begin{center}
\vspace{0.2in}
\includegraphics[width=\columnwidth]{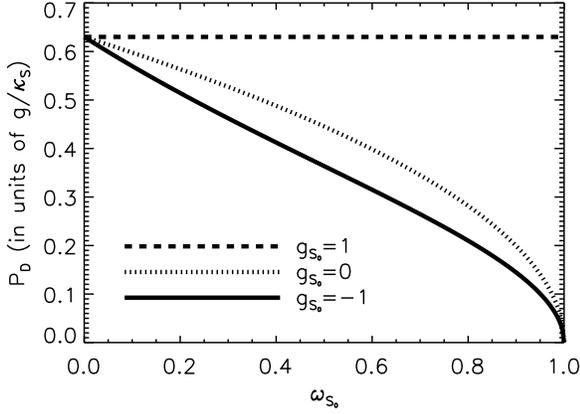}
\end{center}
\vspace{-0.2in}
\caption{Photon deposition depth as a function of the single-scattering albedo, computed for different asymmetry factors and $n=0$.  We have calculated $P_{\rm D}$ in terms of $g/\kappa_{\rm S}$; for example, $g=10^3$ cm s$^{-2}$ and $\kappa_{\rm S} = 0.01$ cm$^2$ g$^{-1}$ yields $g/\kappa_{\rm S} = 0.1$ bar.}
\vspace{0.1in}
\label{fig:pdd}
\end{figure}

\subsection{Longwave}

The longwave refers to the range of wavelengths where the thermal emission of the exoplanet is the dominant source of energy.  It usually occurs in the infrared. 

Integrating the equations in (\ref{eq:tp_govern}) over the longwave, we obtain
\begin{equation}
\begin{split}
\frac{\partial F_{\rm L}}{\partial m} &= \kappa_{\rm L} J_{\rm L} - 4 \kappa_{\rm L}^{\prime\prime} \sigma_{\rm SB} T^4, \\
\frac{\partial K_{\rm L}}{\partial m} &= \frac{ \kappa_{\rm L}^\prime F_{\rm L}}{\beta_{\rm L_0}^2}, \\
\end{split}
\label{eq:longwave_govern}
\end{equation}
where, analogous to the shortwave, we have
\begin{equation}
\beta_{\rm L_0} = \left( \frac{1 - \omega_{\rm L_0}}{1 - \omega_{\rm L_0} g_{\rm L_0}} \right)^{1/2}.
\end{equation}
The absorption, flux and Planck mean opacities are, respectively,
\begin{equation}
\begin{split}
\kappa_{\rm L} &\equiv \frac{\int_{\rm L} \kappa_{\rm a} J ~d\lambda}{\int_{\rm L} J ~d\lambda}, \\
\kappa_{\rm L}^\prime &\equiv \frac{\int_{\rm L} \kappa_{\rm a} F_- ~d\lambda}{\int_{\rm L} F_- ~d\lambda}, \\
\kappa_{\rm L}^{\prime\prime} &\equiv \frac{\pi \int_{\rm L} \kappa_{\rm a} B ~d\lambda}{\sigma_{\rm SB} T^4}. \\
\end{split}
\end{equation}
We assume that $\kappa_{\rm L} = \kappa_{\rm L}^\prime = \kappa_{\rm L}^{\prime\prime}$.  (See also \citealt{hubeny03}.)

\subsection{Derivation of Temperature-Pressure Profile}

Using the second equation in (\ref{eq:longwave_govern}), the first equation in (\ref{eq:energy_conserve_2}) and the $\epsilon_{\rm L_3}$ closure, we obtain
\begin{equation}
J_{\rm L} = J_{\rm L_0} + \frac{1}{\epsilon_{\rm L_3} \beta^2_{\rm L_0}} \int^m_0 \kappa_{\rm L} \left[ {\cal F}_\infty - F_{\rm S} - \tilde{Q}\left(m,\infty\right) \right] ~dm.
\end{equation}
Eliminating the quantities $J_{\rm L}$ and $J_{\rm L_0}$ using equation (\ref{eq:energy_conserve}) and the second equation in (\ref{eq:energy_conserve_2}), respectively, yields
\begin{equation}
\begin{split}
\sigma_{\rm SB} T^4 =& \frac{{\cal F}_\infty}{4} \left( \frac{1}{\epsilon_{\rm L}} + \frac{1}{\epsilon_{\rm L_3} \beta_{\rm L_0}^2} \int^m_0 \kappa_{\rm L} ~dm \right) + {\cal Q} \\
&+ \frac{1}{4} \left( -\frac{F_{\rm S_0}}{\epsilon_{\rm L}} + \frac{\kappa_{\rm S} J_{\rm S}}{\kappa_{\rm L}} - \frac{1}{\epsilon_{\rm L_3} \beta^2_{\rm L_0}} \int^m_0 \kappa_{\rm L} F_{\rm S} ~dm \right). 
\end{split}
\label{eq:tp_intermediate}
\end{equation}
The first term in equation (\ref{eq:tp_intermediate}), associated with ${\cal F}_\infty$, describes the temperature-pressure profile in the deep interior arising from internal heat and is known as ``Milne's solution" \citep{mihalas,mm99}.  It was first derived for stars.  The second term is defined as
\begin{equation}
{\cal Q} \equiv - \frac{1}{4} \left[ \frac{Q}{\kappa_{\rm L}} + \frac{\tilde{Q}\left(0,\infty\right)}{\epsilon_{\rm L}} + \frac{1}{\epsilon_{\rm L_3} \beta^2_{\rm L_0}} \int^m_0 \kappa_{\rm L} \tilde{Q}\left(m,\infty\right) ~dm \right].
\end{equation}
The other terms describe the temperature-pressure profile due to stellar irradiation.  

It is important to note that equation (\ref{eq:tp_intermediate}) does not make any assumptions about the functional forms of $J_{\rm S}$ and $F_{\rm S}$.

The global-mean temperature-pressure profile ($\bar{T}$) is obtained by integrating over $0 \le \phi \le 2\pi$ and $-1 \le \mu \le 0$ and dividing by $2\pi$,
\begin{equation}
\begin{split}
\bar{T}^4 =& \frac{T_{\rm int}^4}{4} \left( \frac{1}{\epsilon_{\rm L}} + \frac{1}{\epsilon_{\rm L_3} \beta^2_{\rm L_0}} \int^m_0 \kappa_{\rm L} ~dm \right) \\
&+ \frac{T_{\rm irr}^4}{8} \left( \frac{1}{2 \epsilon_{\rm L}} + \frac{\kappa_{\rm S} {\cal E}_2}{\kappa_{\rm L} \beta_{\rm S_0}} + \frac{1}{\epsilon_{\rm L_3} \beta^2_{\rm L_0}} \int^m_0 \kappa_{\rm L} {\cal E}_3 ~dm \right) \\
&+ \frac{1}{2\pi} \int^{2\pi}_0 \int^0_{-1} {\cal Q} ~d\mu ~d\phi,
\end{split}
\label{eq:tp_global}
\end{equation}
where ${\cal E}_2 = {\cal E}_2 (\beta_{\rm S})$, ${\cal E}_3 = {\cal E}_3 (\beta_{\rm S})$ and $\beta_{\rm S}$ has previously been defined in equation (\ref{eq:beta_s}).  The factor of 1/8 associated with $T_{\rm irr}^4$ comes about because starlight is incident only upon one hemisphere.

Arguments were previously presented by \cite{guillot10} and \cite{hhps12} for why the last term in equation (\ref{eq:tp_global}) vanishes, based on the reasoning that latitudinal and longitudinal heat transport averages to zero in a global sense.  Within the context of our improved formalism, we find it more natural to simply assert that ${\cal Q} = 0$ when radiative equilibrium is attained ($Q=0$).

\subsection{Temperature-Pressure Profile for a Specific Form of the Longwave Opacity and a Constant Shortwave Opacity}

For equation (\ref{eq:tp_global}) to be useful, we need to explicitly specify the functional form of the longwave opacity,
\begin{equation}
\kappa_{\rm L} = \kappa_0 + \kappa_{\rm CIA} \left( \frac{m}{m_0} \right).
\label{eq:lw_opacity}
\end{equation}
The second term in equation (\ref{eq:lw_opacity}) is used to mimic collision-induced absorption; its associated normalization is $\kappa_{\rm CIA}$.

In radiative equilibrium and for a constant shortwave opacity ($n=0$), combining equations (\ref{eq:tp_global}) and (\ref{eq:lw_opacity}) yields
\begin{equation}
\begin{split}
\bar{T}^4 =& \frac{T_{\rm int}^4}{4} \left[ \frac{1}{\epsilon_{\rm L}} + \frac{m}{\epsilon_{\rm L_3} \beta^2_{\rm L_0}} \left( \kappa_0 + \frac{\kappa_{\rm CIA} m}{2 m_0} \right) \right] \\
&+ \frac{T_{\rm irr}^4}{8} \left[ \frac{1}{2 \epsilon_{\rm L}} + {\cal E}_2 \left( \frac{\kappa_{\rm S}}{\kappa_{\rm L} \beta_{\rm S_0}} - \frac{\kappa_{\rm CIA} m \beta_{\rm S_0}}{\epsilon_{\rm L_3} \kappa_{\rm S} m_0 \beta_{\rm L_0}^2} \right) \right.\\
&\left.+ \frac{\kappa_0\beta_{\rm S_0}}{\epsilon_{\rm L_3} \kappa_{\rm S}\beta_{\rm L_0}^2} \left( \frac{1}{3} - {\cal E}_4 \right) + \frac{\kappa_{\rm CIA} \beta_{\rm S_0}^2}{\epsilon_{\rm L_3} \kappa_{\rm S}^2 m_0 \beta_{\rm L_0}^2} \left( \frac{1}{2} - {\cal E}_3 \right) \right].
\end{split}
\label{eq:tp_global_2}
\end{equation}
As previously mentioned, our formalism yields $\epsilon_{\rm L} = 3/8$ and $\epsilon_{\rm L_3} = 1/3$, but we have intentionally left the values of these Eddington coefficients unspecified in equation (\ref{eq:tp_global_2}) to allow for other choices to be made, if desired.

\section{Application to Exoplanets}
\label{sect:apps}

\begin{table*}
\label{tab:closures2}
\begin{center}
\caption{Various Choices for Coefficients of Two-Stream Equations (Closures)}
\begin{tabular}{lcccccc}
\hline
\hline
Name & $\gamma_{\rm a}/\bar{\mu}$ & $\gamma_{\rm s}/\bar{\mu}$ & $\gamma_{\rm B}/\bar{\mu}$ & $f_\infty$ & $\left( \gamma_{\rm a} - \gamma_{\rm s} \right)/\bar{\mu}$ & References \\
\hline
Hemispheric / hemi-isotropic & $2 - \omega_0 \left( 1 + g_0 \right)$ & $\omega_0 \left( 1 - g_0 \right)$ & $2\pi\left(1-\omega_0\right)$ & 1 & $2\left( 1 - \omega_0 \right)$ & MW80, T89, P10, HML \\
Eddington & $\frac{1}{4} \left[ 7 - \omega_0 \left(4 + 3 g_0 \right) \right]$ & $-\frac{1}{4} \left[ 1 - \omega_0 \left( 4 - 3 g_0 \right) \right]$ & $2\pi\left(1-\omega_0\right)$ & $\gtrsim 1.8$ & $1 - \omega_0$ & MW80, GY89, T89, P10 \\
Quadrature & $\frac{\sqrt{3}}{2} \left[ 2 - \omega_0 \left( 1+g_0 \right) \right]$ & $\frac{\sqrt{3}\omega_0}{2} \left( 1 - g_0 \right)$ & $\sqrt{3}\pi\left(1-\omega_0\right)$ & 1 & $\sqrt{3} \left( 1 - \omega_0 \right)$ &  MW80, T89, P10 \\
\hline
\hline
\end{tabular}\\
MW80: \cite{mw80}, GY89: \cite{gy89}, T89: \cite{toon89}, P10: \cite{pierrehumbert}, HML: this study.\\
\end{center}
\end{table*}

\subsection{Other Closures for Two-Stream Radiative Transfer: Comparison and Implications}
\label{subsect:other_closures}

\begin{figure}
\begin{center}
\includegraphics[width=\columnwidth]{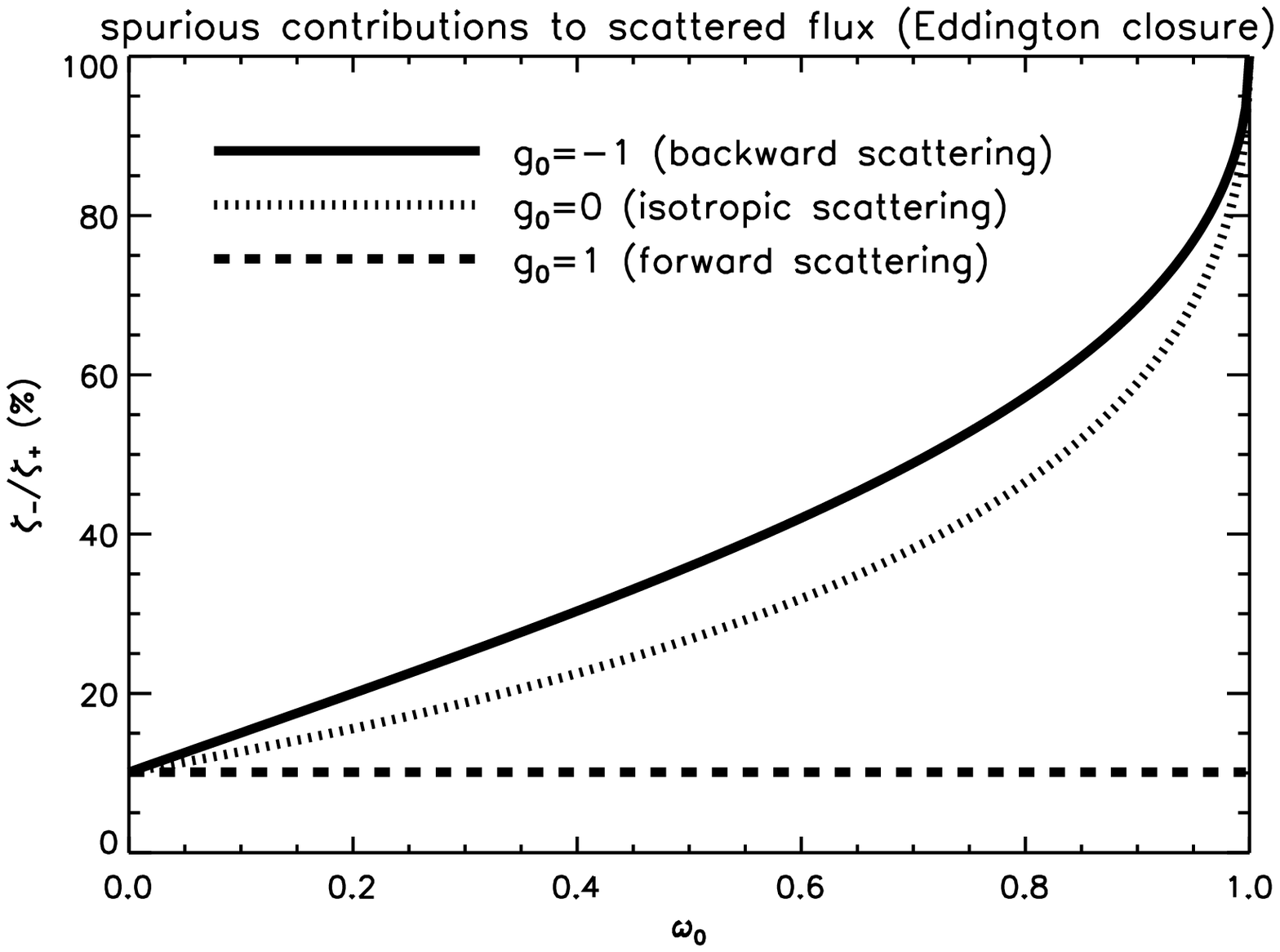}
\includegraphics[width=\columnwidth]{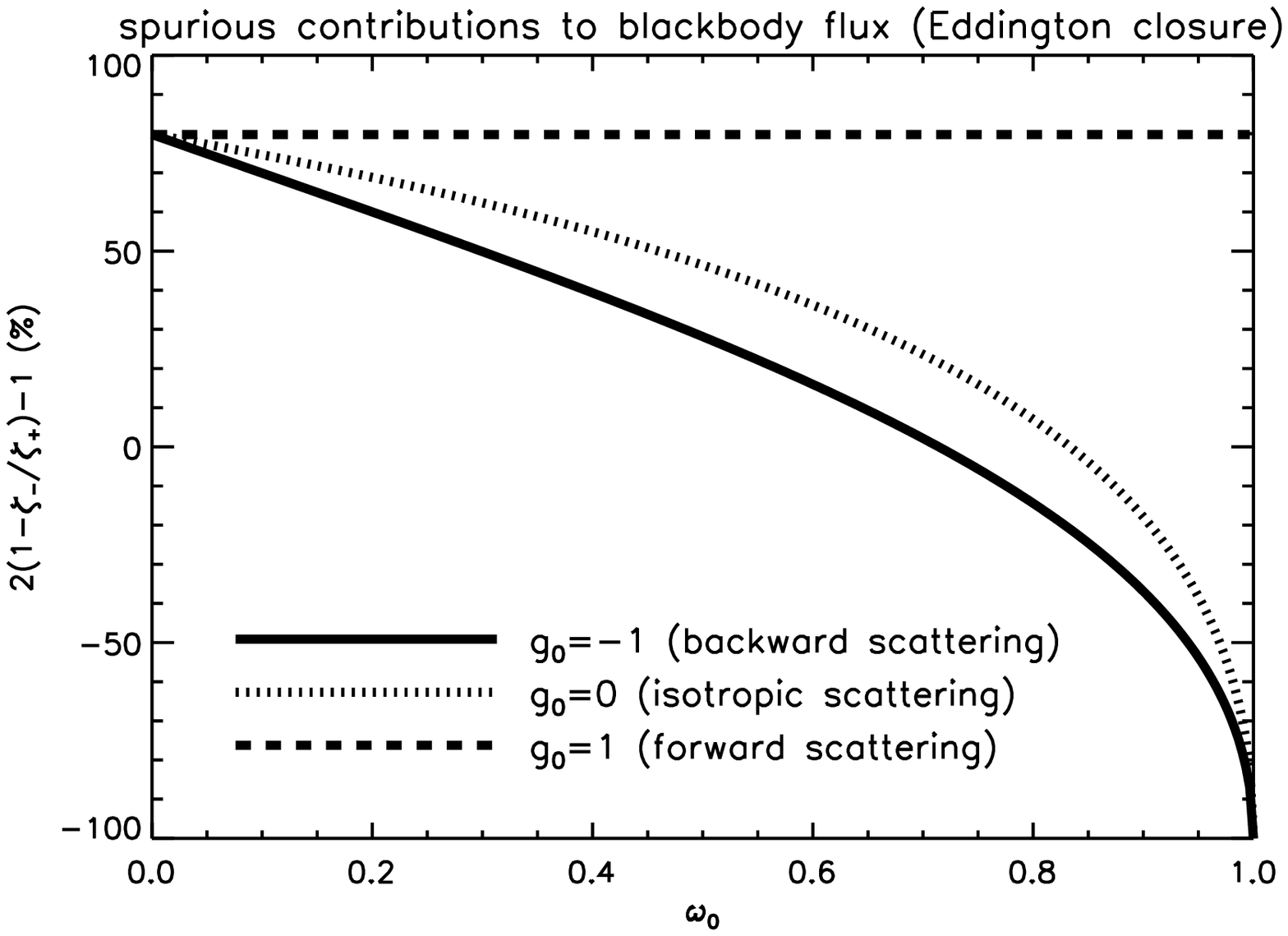}
\end{center}
\vspace{-0.2in}
\caption{Errors incurred when using the Eddington closure for the two-stream approximation in the limit of non-isotropic, coherent scattering.  Top panel: error expressed as a percentage of the reflected flux.  Bottom panel: error expressed as a spurious percentage enhancement of the blackbody flux.}
\label{fig:eddington}
\end{figure}

There is a rich literature describing various forms of the two-stream equations \citep{chandra60,mihalas,mw80,gy89,toon89,mm99}.  Specifically, it boils down to having different expressions for the coefficients $\gamma_{\rm a}$, $\gamma_{\rm s}$ and $\gamma_{\rm B}$ in the two-stream equations for the outgoing and incoming fluxes, which in turn depends on the choice of closures (the Eddington coefficients).  In this subsection, we will explore these other choices published in the literature and examine their implications.

Table 3 lists the choices of $\gamma_{\rm a}$, $\gamma_{\rm s}$ and $\gamma_{\rm B}$ for different closures.  We do not discuss closures that involve a series expansion of the scattering phase function in terms of Legendre polynomials \citep{chandra60,mw80}.  Energy conservation in the purely scattering limit requires that we check the $\left( \gamma_{\rm a} - \gamma_{\rm s} \right)$ expressions for each closure (see \S\ref{subsect:enforce_scat}).  Since the hemispheric/hemi-isotropic, Eddington and quadrature closures all have $\left( \gamma_{\rm a} - \gamma_{\rm s} \right)\propto \left( 1 - \omega_0 \right)$, they all ensure that radiative equilibrium is attained ($\partial F_-/\partial \tau = 0$) when $\omega_0=1$.

Next, we need to check the total flux in the limit of a purely absorbing, opaque atmosphere.  For the hemispheric/hemi-isotropic and quadrature closures, we have $\gamma_{\rm s}=0$ when $\omega_0=0$, which implies that $\zeta_+=1$ and $\zeta_-=0$.  For these closures, we may easily define the dimensionless factor,
\begin{equation}
f_\infty \equiv \frac{\gamma_{\rm B}}{\pi \left( \gamma_{\rm a} - \gamma_{\rm s} \right)},
\end{equation}
which is the limiting value of $F_+$ as ${\cal T} \rightarrow 0$, normalized by $2\pi B$.  We verify that $f_\infty=1$ for the hemispheric/hemi-isotropic and quadrature closures.  

For the Eddington closure, the issue is more subtle.  At first glance, one may already anticipate that the Eddington closure is unphysical, as $\gamma_{\rm s} \ne 0$ even in the absence of scattering---somehow, some fraction of the outgoing rays still gets converted into incoming ones (and vice versa).  The total flux has a limiting value as ${\cal T} \rightarrow 0$,
\begin{equation}
F_+ \rightarrow \frac{\zeta_-}{\zeta_+} \left( F_{\downarrow_1} + F_{\uparrow_2} \right) + \frac{2\gamma_{\rm B} B}{\gamma_{\rm a} - \gamma_{\rm s}} \left( 1 - \frac{\zeta_-}{\zeta_+} \right).
\label{eq:flux_limit_eddington}
\end{equation}
Unlike for the other closures, we have $\zeta_- \ne 0$ even when $\omega_0=0$.  Specifically, we have
\begin{equation}
\frac{\zeta_-}{\zeta_+} = \frac{1 - \left[2 \left( 1 - \omega_0 \right)/3\left( 1 - \omega_0 g_0 \right) \right]^{1/2}}{1 + \left[2 \left( 1 - \omega_0 \right)/3\left( 1 - \omega_0 g_0 \right) \right]^{1/2}}.
\end{equation}
It is also worth noting that the boundary condition $F_{\downarrow_1}$ is associated with $F_{\uparrow_1}$, while $F_{\uparrow_2}$ is associated with $F_{\downarrow_2}$, implying that the limiting values of the  incoming and outgoing fluxes behave as if reflection is present (in the form of $\sim 10$--100\% contributions from the boundary conditions), even in the purely absorbing limit.  Such contributions are unphysical.  Thus, using the Eddington closure leads to two types of error: a spurious contribution due to reflected flux and a spurious enhancement of the blackbody flux.  (See also the caption of Figure 3 of \citealt{toon89} and their summary section.)  Figure \ref{fig:eddington} shows the percentage errors, associated with both artifacts, incurred when using the Eddington closure.  It is apparent that when $g_0 \ne 1$, the errors are non-uniform as they depend both on $g_0$ and $\omega_0$.  Without knowing what these boundary conditions generally are, we may set a lower limit to $f_\infty$ by considering the term associated with $B$ in equation (\ref{eq:flux_limit_eddington}),
\begin{equation}
f_\infty \ge \frac{\gamma_{\rm B}}{\pi \left( \gamma_{\rm a} - \gamma_{\rm s}\right) } \left( 1 - \frac{\zeta_-}{\zeta_+} \right).
\end{equation}
For $\omega_0=0$, we have $\zeta_-/\zeta_+ = 5 - 2\sqrt{6} \approx 0.1$.  Since $\gamma_{\rm B}/( \gamma_{\rm a} - \gamma_{\rm s} ) = 2 \pi$ for the Eddington closure, we obtain $f_\infty \gtrsim 1.8$.  Overall, we recommend that the Eddington closure not be used as it produces spurious reflected fluxes, artificially enhances the blackbody flux and the associated errors are non-uniform (and therefore challenging to quantify between different model atmospheres).

\begin{figure}
\begin{center}
\includegraphics[width=\columnwidth]{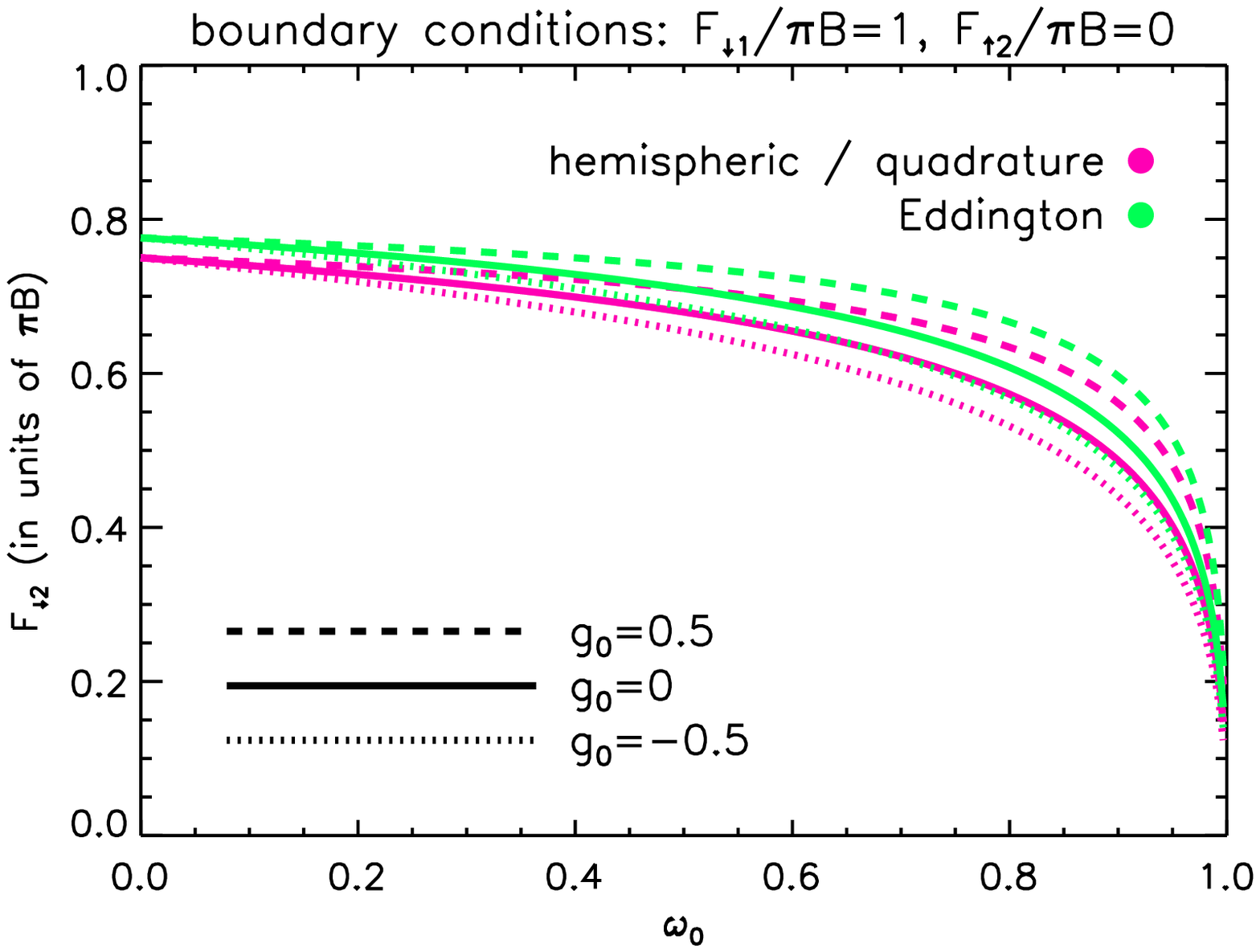}
\includegraphics[width=\columnwidth]{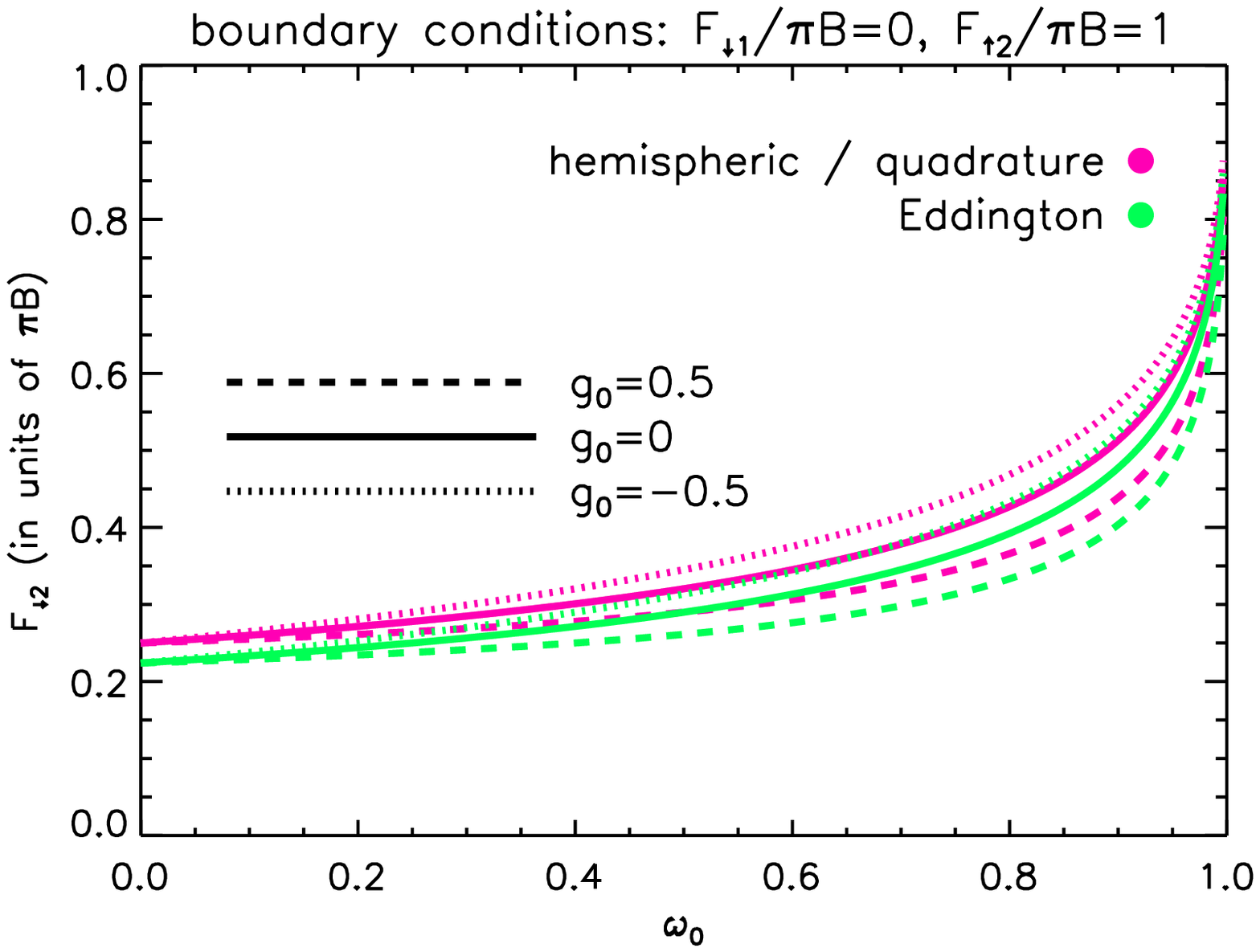}
\end{center}
\vspace{-0.2in}
\caption{Incoming flux as a function of the single-scattering albedo ($\omega_0$) for coherent, non-isotropic scattering.  We have adopted ${\cal T}=0.5$ for illustration.  All fluxes and their boundary conditions are given in terms of the blackbody flux ($\pi B$).}
\label{fig:fluxes_non_iso}
\end{figure}

Figure \ref{fig:fluxes_non_iso} shows examples of the incoming flux ($F_{\downarrow_2}$) as a function of $\omega_0$ for different values of $g_0$.  We show only the incoming flux as its expression is identical to that for the outgoing flux except for the boundary conditions.  Since the hemispheric and quadrature closures yield identical expressions for $\zeta_\pm$ and $\gamma_{\rm B}/(\gamma_{\rm a} - \gamma_{\rm s})$, they produce identical fluxes.  With our chosen boundary conditions ($F_{\downarrow_1}/\pi B = 1$, $F_{\uparrow_2}/\pi B = 0$), adopting the Eddington closure results in an over-estimation of the incoming fluxes.  If we reverse the boundary conditions ($F_{\downarrow_1}/\pi B = 0$, $F_{\uparrow_2}/\pi B = 1$), the Eddington closure now produces an under-estimation of the incoming fluxes.  The errors are non-uniform (unless $g_0=1$) and typically $\sim 1$--10\%, depending on the boundary conditions adopted as well as the values of ${\cal T}$, $\omega_0$ and $g_0$.

We conclude that, for the purpose of exploring parameter space and studying trends associated with exoplanetary atmospheres, the two-stream radiative transfer treatment should only be used with the hemispheric or hemi-isotropic closure.  The Eddington closure should be avoided.

\subsection{Recipes for Applying the Hybrid Technique of Two-Stream and Flux-Limited-Diffusion Radiative Transfer Towards Computing Synthetic Spectra}
\label{subsect:recipe}

We concoct computational recipes for calculating synthetic spectra using the two-stream radiative transfer, augmented by flux-limited diffusion in the deep atmosphere.  The first recipe describes a stand-alone, one-dimensional (1D) calculation ignoring the effects of atmospheric dynamics.  The second recipe describes how to couple the radiative transfer scheme to a three-dimensional (3D) general circulation model of the atmosphere.

\subsubsection{1D Purely Radiative Atmosphere}

\begin{enumerate}

\item Specify an initial guess for the temperature-pressure profile.  Specify the boundary conditions at the bottom (internal heat) and top (stellar irradiation) of the atmosphere.  For the former, it is important to note that the \textit{net flux} is $\pi B(T_{\rm int})$; the outgoing flux at the bottom of the atmosphere is then an iterative boundary condition.

\item Use the equations in (\ref{eq:twostream_iso_general_solution_non_isothermal_2}) or (\ref{eq:twostream_iso_general_solution_non_isothermal}) to perform the two-stream calculation by populating each layer of the model atmosphere with outgoing and incoming fluxes.

\item Integrate the net flux over wavelength, compute its gradient and update the temperature-pressure profile using
\begin{equation}
T_{\rm new}  = T_{\rm old} + \frac{\Delta t}{\rho c_P} \frac{\partial {\cal F}_-}{\partial z},
\label{eq:heat_equation0}
\end{equation}
where $\Delta t$ is the computational time step and $c_P$ is the specific heat capacity at constant pressure.  Note that the vertical coordinate ($z$) is defined from the top of the atmosphere downwards.

\item Repeat steps 1 to 3 until radiative equilibrium is attained ($\partial {\cal F}_-/\partial z = 0$).  The synthetic spectrum is given by $F_\uparrow$, across wavelength, at the top of the computational domain.

\end{enumerate}

For exoplanets with surfaces, the flux from the surface is specified as the bottom boundary condition.  For gaseous exoplanets with $T_{\rm irr} \gg T_{\rm int}$, the two-stream recipe is first implemented with $T_{\rm int}=0$, after which the interior temperature-pressure profile is added using the flux-limited-diffusion solution in equation (\ref{eq:tdeep}).\footnote{Note that one adds the fluxes and not the temperatures.}  Thus, heating in the deep interior is performed semi-analytically; by definition, the solution in equation (\ref{eq:tdeep}) is in radiative equilibrium.  One may need a convective adjustment scheme for treating convectively unstable parts of the temperature-pressure profile \citep{manabe65}.

\subsubsection{3D Radiative Atmosphere with Dynamics}

\begin{enumerate}

\item Instead of iterating for radiative equilibrium within the 1D radiative transfer solver, compute the wavelength-integrated net flux (${\cal F}_-$) and feed it to a more general expression for the heat equation, which we will now derive.  The first law of thermodynamics states,
\begin{equation}
Q = c_V \frac{DT}{Dt} + P \frac{DV}{Dt},
\end{equation}
where $Q$ represents all forms of heating, $c_V$ is the specific heat at constant volume and $V = 1/\rho$ is the specific volume.  Using the ideal gas law ($P = \rho {\cal R} T$, where ${\cal R}$ is the specific gas constant) and $c_P = c_V + {\cal R}$, we obtain
\begin{equation}
\rho c_P \frac{DT}{Dt} = \rho Q + \frac{DP}{Dt}.
\label{eq:heat_equation}
\end{equation}
If we ignore conduction, then the energy per unit volume and time associated with heating is
\begin{equation}
\rho Q = - \nabla. \vec{{\cal F}}_- = \frac{\partial {\cal F}_-}{\partial z}.
\end{equation}
Equation (\ref{eq:heat_equation}) is solved in tandem with the Navier-Stokes and mass continuity equations to self-consistently obtain $T$, $\rho$ and $\vec{v}$ (the velocity field).

\item The new temperature-pressure profile (iterated consistently with the velocity field) is fed back to the 1D radiative transfer solver to obtain updated values of ${\cal F}_-$.  The entire process is repeated until the simulation reaches equilibrium.

\end{enumerate}

In the absence of atmospheric dynamics, we may write $DT/Dt \equiv \partial T/ \partial t + \vec{v}.\nabla T \approx \partial T/ \partial t$ and ignore ``$PdV$" work (i.e., set $DP/Dt = 0$).  Under such restricted conditions, we obtain equation (\ref{eq:heat_equation0}).  

Under terrestrial conditions, we may safely assume that $\partial T/\partial t \gg \vec{v}.\nabla T$.  At the order-of-magnitude level, the terms are
\begin{equation}
\frac{\partial T}{\partial t} \sim \frac{T}{t_{\rm rad}} = \frac{g \sigma_{\rm SB} T^4}{c_P P}
\end{equation}
and 
\begin{equation}
\vec{v}.\nabla T \sim \frac{v_\phi T}{R},
\end{equation}
where $v_\phi$ is the zonal velocity and $R$ is the radius of the exoplanet.  For highly-irradiated atmospheres, the advection term cannot be ignored when
\begin{equation}
\begin{split}
P >& 0.06 \mbox{ bar} ~\left( \frac{g}{10 \mbox{ m s}^{-2}} \frac{R}{10^{10} \mbox{ cm}} \right) \left( \frac{T}{10^3 \mbox{ K}} \right)^3 \\
&\times \left( \frac{v_\phi}{1 \mbox{ km s}^{-1}} \frac{c_P}{10^8 \mbox{ erg K}^{-1} \mbox{ g}^{-1}} \right)^{-1}.
\end{split}
\end{equation}

\subsection{The Bond, Spherical and Geometric Albedos and Albedo Spectra}

\begin{figure}
\begin{center}
\vspace{0.2in}
\includegraphics[width=\columnwidth]{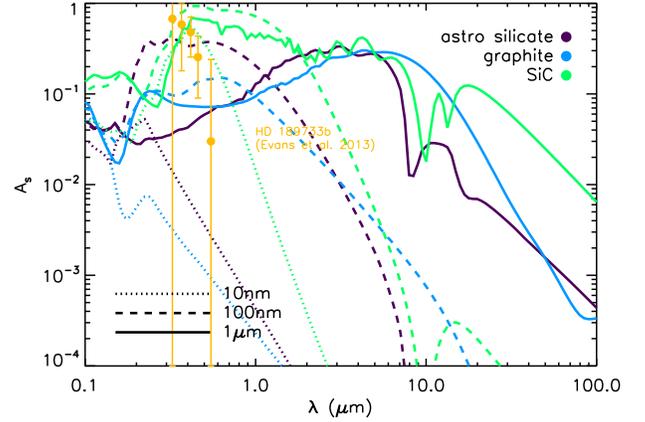}
\end{center}
\vspace{-0.2in}
\caption{Spherical albedo versus wavelength for dust grains of different radii and compositions.  No extra optical absorber (e.g., sodium atoms) is included.}
\vspace{0.1in}
\label{fig:asphere}
\end{figure}

The spherical albedo ($A_{\rm s}$) is the ratio of scattered to incident flux (e.g., \citealt{seager10}).  The Bond albedo ($A_{\rm B}$) is the spherical albedo integrated over all wavelengths.  Within the context of the two-stream approximation, the expressions for both quantities are presented in equation (\ref{eq:bond_albedo}).  

However, secondary eclipse measurements in the optical measure the geometric albedo ($A_g$), assuming that thermal emission from the exoplanet does not contaminate the signal \citep{hd13,angerhausen14}.  To convert between $A_{\rm s}$ and $A_g$ requires knowledge of the \textit{scattered} flux at all phase angles, which is beyond the scope of our current two-stream treatment.  Specifically, one needs to evaluate the phase integral \citep{russell16,marley99,seager10,mb12},
\begin{equation}
q = 2 \int^\pi_0 \frac{F_{\rm scat}}{F_{\rm scat,0}} \sin\psi ~d\psi,
\end{equation}
where $\psi$ is the phase angle, $F_{\rm scat}(\psi)$ is the emergent scattered flux and $F_{\rm scat,0} \equiv F_{\rm scat}(\psi=0)$.  The geometric albedo is defined at zero phase angle.  The spherical and geometric albedos are related by $A_{\rm s} = q A_g$.  For a Lambert sphere (isotropic scattering), we have $A_g = 2 A_{\rm s}/3$.  For Rayleigh scattering, $A_g = 3 A_{\rm s}/4$.  Generally, the conversion factor between the spherical and geometric albedos is an order-of-unity constant for a specific scattering profile and at a given wavelength.

In Figure \ref{fig:asphere}, we show examples of $A_{\rm s}$ for dust grains composed of astronomical silicate, graphite and silicon carbide (SiC), where the tabulated data for $\omega_0$ and $g_0$ have been taken from the full Mie calculations of \cite{draine84} and \cite{laor93}.  We have not included extra sources of absorption (e.g., sodium atoms), unlike in \cite{hd13}.  We include the measured values of the geometric albedo of HD 189733b, by \cite{evans13}, and assume $A_{\rm s} = 3A_g/2$ for these data points.  As expected, there is a strong dependence of $A_{\rm s}$ on the dust grain radius and somewhat less on the composition \citep{pierrehumbert,hd13}.  Curiously, the measured albedo spectrum of HD 189733b is consistent with an atmosphere populated by silicon carbide grains, with radii of 10 nm, without a need for an extra optical absorber.

\subsection{Analytical Temperature-Pressure Profiles with Non-Isotropic Scattering}
\label{subsect:tp_results}

\begin{figure}
\begin{center}
\vspace{0.2in}
\includegraphics[width=\columnwidth]{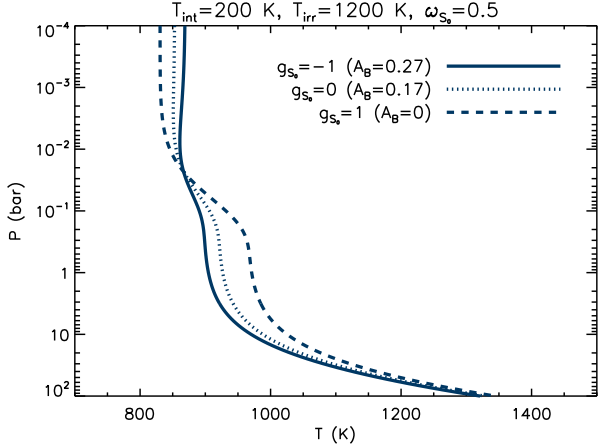}
\end{center}
\vspace{-0.2in}
\caption{Temperature-pressure profiles for different values of the asymmetry factor in the optical or shortwave and a constant shortwave opacity ($n=0$).}
\vspace{0.1in}
\label{fig:tp1}
\end{figure}

\begin{figure}
\begin{center}
\vspace{0.2in}
\includegraphics[width=\columnwidth]{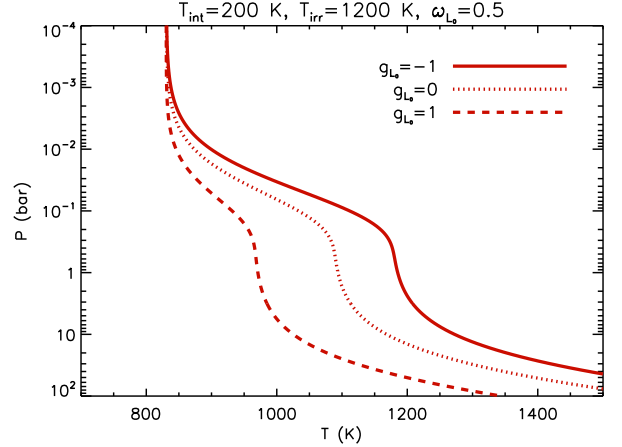}
\end{center}
\vspace{-0.2in}
\caption{Temperature-pressure profiles for different values of the asymmetry factor in the infrared or longwave and a constant shortwave opacity ($n=0$).}
\vspace{0.1in}
\label{fig:tp2}
\end{figure}

We now elucidate trends in the temperature-pressure profiles with non-isotropic scattering, building on the work of \cite{guillot10} (pure absorption) and \cite{hhps12} (isotropic scattering).  In Figure \ref{fig:tp1}, we first examine the effects of varying the asymmetry factor in the optical or shortwave ($g_{\rm S_0}$).  Physically, this has a couple of effects: altering the Bond albedo and changing the location of the photon deposition layer.  As an illustration, we assume a constant shortwave opacity ($n=0$) and use equation (\ref{eq:tp_global_2}).  We fix $\omega_{\rm S_0}=0.5$, which yields $\beta_{\rm S_0} = 1/\sqrt{3}, 1/\sqrt{2}$ and 1 for $g_{\rm S_0} = -1, 0$ and 1, respectively.  Correspondingly, we have $A_{\rm B} \approx 0.27, 0.17$ and $=0$ and $P_{\rm D} \approx 36, 46$ and $=63$ mbar.  We set $\kappa_{\rm S} = 0.01$ cm$^2$ g$^{-1}$, $\kappa_0 = 0.02$ cm$^2$ g$^{-1}$, $\kappa_{\rm CIA}=0$, $g= 10^3$ cm s$^{-2}$, $T_{\rm int}=200$ K and $T_{\rm irr}=1200$ K.  Although the temperature-pressure profile with backward scattering ($g_{\rm S_0}=-1$) is mostly cooler than the profiles with isotropic and forward scattering, it is warmer at low pressures due to the photon deposition depth being located at a higher altitude.  Non-isotropic scattering introduces an anti-greenhouse effect as scattering becomes more backward-peaked.

In Figure \ref{fig:tp2}, we assume $A_{\rm B}=0$ and examine the effects of varying the asymmetry factor in the infrared or longwave ($g_{\rm L_0}$).  Any form of infrared scattering generally warms the atmosphere, unless it takes the form of purely forward scattering, which behaves like pure absorption---the ``scattering greenhouse effect" \citep{pierrehumbert}.

\begin{figure}
\begin{center}
\vspace{0.2in}
\includegraphics[width=\columnwidth]{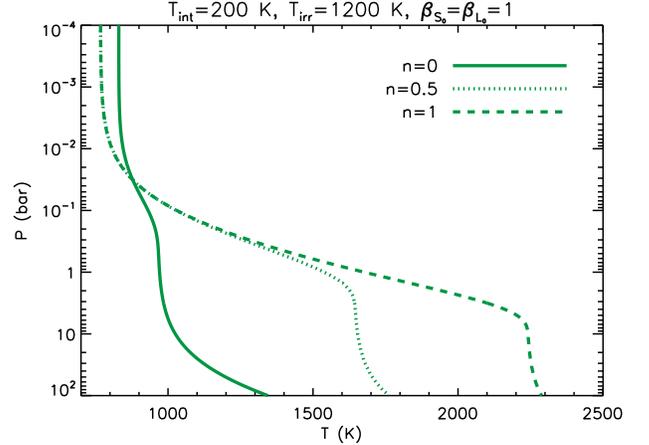}
\end{center}
\vspace{-0.2in}
\caption{Temperature-pressure profiles with constant ($n=0$) and non-constant ($n \ne 0$) optical or shortwave opacities.}
\vspace{0.1in}
\label{fig:tp3}
\end{figure}

To investigate the effects of a non-constant optical/shortwave opacity, we evaluate equation (\ref{eq:tp_global}) numerically\footnote{The need to compute some of these terms numerically is the basis for \cite{hhps12} describing these models as being ``semi-analytical".} and in radiative equilibrium.  Figure \ref{fig:tp3} shows examples of temperature-pressure profiles with $n=0, 0.5$ and 1.  The photon deposition depth resides deeper as $n$ increases, leading to warmer profiles at higher pressures.  The model atmosphere generally becomes less isothermal, which is partially an artifact of assuming a constant optical opacity.

\subsection{Is the Photosphere Always at an Optical Depth of 2/3?}

Radiation is typically absorbed, scattered or emitted at optical depths $\sim 1$.  In self-luminous atmospheres (e.g., stars), the classical Milne's solution is \citep{mihalas,mm99}
\begin{equation}
\bar{T} = T_{\rm int} \left[ \frac{3}{4} \left( \frac{2}{3} + \tau_{\rm L} \right) \right]^{1/4},
\end{equation}
where $\tau_{\rm L}$ is the infrared or longwave optical depth.  When $\tau_{\rm L}=2/3$, we have $\bar{T} = T_{\rm int}$.  In stars, $T_{\rm int} = T_\star$.  This is the basis for stating that the solar photosphere occurs at an optical depth of 2/3, where we sample $T_\star \approx 5800$ K (instead of either the $\sim 10^4$ K chromosphere at $\tau_{\rm L} \ll 1$ or the $\sim 10^7$ K deep interior of the Sun at $\tau_{\rm L} \gg 1$).

With our choice of closures (Table 2), Milne's solution is generalized to
\begin{equation}
\bar{T} = T_{\rm int} \left\{ \frac{3}{4} \left[ \frac{8}{9} + \left( 1 - \omega_{\rm L_0} g_{\rm L_0} \right) \tau_{\rm L} \right] \right\}^{1/4},
\end{equation}
where we have defined $\tau_{\rm L} \equiv \int^m_0 \kappa_{\rm L} dm / ( 1 - \omega_{\rm L_0} )$.  In the presence of scattering, the photosphere for self-luminous objects resides at an optical depth of
\begin{equation}
\tau_{\rm L} = \frac{4}{9 \left( 1 - \omega_{\rm L_0} g_{\rm L_0} \right)}.
\label{eq:milne}
\end{equation}

For atmospheres with both stellar irradiation and internal heat, one has to obey energy conservation by setting $\bar{T}^4 = T_{\rm int}^4 + T_{\rm irr}^4/4$ in equation (\ref{eq:tp_global}) and solving for $\tau_{\rm L}$.\footnote{A factor of 1/2 comes from considering stellar irradiation onto one hemisphere only, while the other factor of 1/2 comes from averaging over $\mu F_\star$.}  For fixed values of the optical/shortwave and infrared/longwave opacities, $\tau_{\rm L}$ is independent of the value of $T_{\rm irr}$.

In Figure \ref{fig:tau}, we show calculations of $\tau_{\rm L}$ as a function of $\kappa_0$ for $g_{\rm L_0} = -1, 0$ and 1, using the values of the parameters stated in \S\ref{subsect:tp_results}.  Consistent with the temperature-pressure profiles showed in Figure \ref{fig:tp2}, the infrared photosphere resides higher up in the atmosphere as longwave scattering becomes more backward-peaked.  The dependence on the infrared opacity is generally weak.

\begin{figure}
\begin{center}
\vspace{0.2in}
\includegraphics[width=\columnwidth]{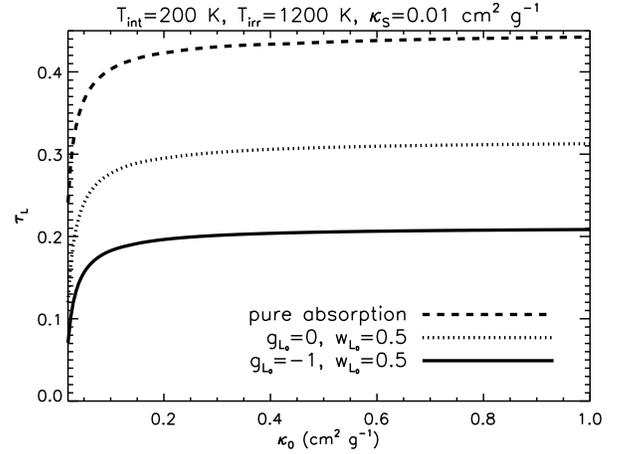}
\end{center}
\vspace{-0.2in}
\caption{Infrared photosphere as a function of the infrared opacity.  For illustration, we have assumed pure absorption in the optical or shortwave ($\beta_{\rm S_0}=0$) and a constant shortwave opacity ($n=0$).}
\vspace{0.1in}
\label{fig:tau}
\end{figure}

\subsection{Toy Models of the Runaway Greenhouse Effect: the Komabayashi-Ingersoll Limit}

As an illustration of the versatility of our two-stream solutions, we now use them to construct toy models of the runaway greenhouse effect \citep{koma67,ingersoll69}.  Consider an atmosphere with a single condensible component, initially existing in liquid or solid form.  As the surface temperature rises, it is transformed into its gaseous form, which triggers a positive feedback reaction where warmer temperatures produce even more warming by releasing more greenhouse gas.  The atmosphere attempts to cool itself by increasing its ``outgoing longwave radiation" (OLR), which is the emergent infrared flux.  The essence of the runaway greenhouse effect is that there is a limit to which the OLR may increase, due to the transmissivity of the atmosphere falling to zero.  \cite{pierrehumbert} calls this the ``Komabayashi-Ingersoll limit". 

We use equation (\ref{eq:twostream_iso_general_solution_non_isothermal}) with ${\cal F}_{\rm OLR} = \int_{\rm L} F_{\uparrow_1} d\lambda$, $\int_{\rm L} F_{\downarrow_1} d\lambda=0$ (negligible starlight in the infrared) and $\int_{\rm L} F_{\uparrow_2} d\lambda = \sigma_{\rm SB} T_{\rm s}^4$ with $T_{\rm s}$ being the surface temperature.  The transmission function is
\begin{equation}
{\cal T} = \exp{\left(-\tau_{\rm s} \right)},
\end{equation}
with the total optical depth of the atmosphere being
\begin{equation}
\tau_{\rm s} = \frac{\kappa_{\rm L} P_{\rm s}}{g \left( 1 - \omega_{\rm L_0} \right)},
\label{eq:tau_s}
\end{equation}
and $P_{\rm s}$ being the surface pressure.  We will assume that the atmosphere is saturated, such that the temperature and pressure are related by the Clausius-Clapeyron equation,
\begin{equation}
P = P_{\rm cc} \exp{\left( - \frac{T_{\rm cc}}{T} \right)},
\label{eq:cc}
\end{equation}
where $P_{\rm cc}$ and $T_{\rm cc}$ are normalizations for the pressure and temperature, respectively.  This approximate form of the Clausius-Clapeyron equation assumes a constant specific latent heat of condensation or sublimation with temperature; values of $P_{\rm cc}$ and $T_{\rm cc}$ may be found in Table 1 of \cite{hk12}.  Equation (\ref{eq:cc}) may be used to relate $T_{\rm s}$ and $P_{\rm s}$.  It is also used to compute the temperature in $\int_{\rm L} \pi B d\lambda = \sigma_{\rm SB} T^4$ with the pressure now being given by the photospheric pressure, $P = 4 g \beta_{\rm L_0}^2/9\kappa_{\rm L}$, via use of equation (\ref{eq:milne}).

Figure \ref{fig:olr} shows calculations of the OLR flux (${\cal F}_{\rm OLR}$) for atmospheres containing only water, ammonia, carbon dioxide or methane.  For illustration, we have chosen $g=10^3$ cm s$^{-2}$ and $\kappa_{\rm L} = 10^{-6}$ cm$^2$ g$^{-1}$.   At low surface temperatures (and $\tau_{\rm s}$), we have ${\cal F}_{\rm OLR} \approx \sigma_{\rm SB} T^4_{\rm s}$.  As the surface temperature rises to the point where ${\cal T}=0$, the OLR asymptotes to $\int_{\rm L} \pi B d\lambda$.  The non-monotonic behavior of ${\cal F}_{\rm OLR}$, as it transitions between the two regimes, is an artifact of using an isothermal solution to approximate non-isothermal behavior.  As expected, the presence of infrared scattering ($g_{\rm L_0} < 1$) results in warmer atmospheres and a lower value of the Komabayashi-Ingersoll limit, implying that the runaway greenhouse is more easily triggered.

The conservation of energy dictates that the OLR flux needs to be equal to the incoming stellar flux: ${\cal F}_{\rm OLR} = {\cal L}_\star/4\pi a^2$, where ${\cal L}_\star$ is the stellar luminosity.  Denoting the stellar mass by $M_\star$, one may obtain the inner boundary of the habitable zone by using the appropriate ${\cal L}_\star (M_\star)$ relationship for stars.  Our value of $\kappa_{\rm L}$ was chosen such that, for a Sun-like star, $a \approx 0.7$ AU for a purely absorbing atmosphere.  Such a freedom to specify $\kappa_{\rm L}$ reflects the inability of our toy models to make quantitative predictions for the runaway greenhouse effect, a property already noted by \cite{pierrehumbert}, although they provide useful tools for understanding basic trends.

\begin{figure}
\begin{center}
\vspace{0.2in}
\includegraphics[width=\columnwidth]{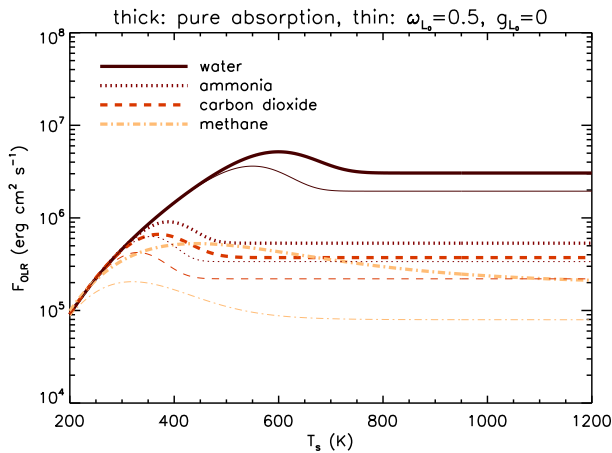}
\end{center}
\vspace{-0.2in}
\caption{Outgoing longwave radiation as a function of the surface temperature of the exoplanet for different greenhouse gases.}
\vspace{0.1in}
\label{fig:olr}
\end{figure}

\section{Discussion}
\label{sect:discussion}

\subsection{Summary}

The salient points of our study may be summarized as follows.
\begin{itemize}

\item \textbf{Unified, self-consistent framework: } Starting from the radiative transfer equation, we have derived a set of governing equations and solutions that include the two-stream treatment, a transition to the diffusion approximation in the deep interior of the exoplanet and temperature-pressure profiles involving non-isotropic scattering.  Utilizing these solutions requires a set of closures (Eddington coefficients) to be specified, which we derive self-consistently based on energy conservation.

\item \textbf{Framework for computing synthetic spectra: } From the two-stream equations, we derived the outgoing and incoming fluxes as functions of wavelength, single-scattering albedo, asymmetry factor and transmission function, as stated in equations (\ref{eq:twostream_iso_general_solution_non_isothermal_2}) and  (\ref{eq:twostream_iso_general_solution_non_isothermal}) for non-isothermal and isothermal model layers, respectively.  The outgoing flux as a function of wavelength is the synthetic spectrum.  In \S\ref{subsect:recipe}, we have provided recipes for using these solutions either in stand-alone calculations of synthetic spectra or general circulation models.

\item \textbf{Temperature-pressure profiles with non-isotropic scattering: } Using the dual-band approximation, we have derived analytical solutions, in equation (\ref{eq:tp_global_2}), for the temperature-pressure profiles in the limit of non-isotropic, coherent scattering and a constant optical opacity.  For a non-constant (power-law) optical opacity, one has to use equation (\ref{eq:tp_global}).  As scattering in the optical becomes more backward-peaked, it introduces an anti-greenhouse effect to the thermal structure.  Scattering in the infrared generally warms the atmosphere (i.e., the scattering greenhouse effect), unless it is in the form of purely forward scattering, in which case it behaves like a purely absorbing atmosphere.

\item \textbf{Spherical and Bond albedos and albedo spectra: } We have derived analytical formulae for the spherical and Bond albedos, in the two-stream approximation, as functions of the single-scattering albedo ($\omega_0$) and asymmetry factor ($g_0$), as stated in equation (\ref{eq:bond_albedo}).  The formula for the spherical albedo may be used to compute albedo spectra using tabulated values of $\omega_0$ and $g_0$, examples of which are shown in Figure \ref{fig:asphere}.

\item \textbf{Photon deposition depth with non-isotropic scattering: } The photon deposition depth is the pressure at which most of the incident stellar irradiation is being absorbed.  In equation (\ref{eq:pdd}), we derive an updated formula that involves non-isotropic scattering and a non-constant optical opacity.

\item \textbf{Runaway greenhouse: } Our two-stream solution allows us to construct toy models for the runaway greenhouse effect and compute the Komabayashi-Ingersoll limit for different gases.

\item \textbf{Use the hemispheric closure, avoid the Eddington closure: } The Eddington closure leads to two forms of error.  First, it introduces reflected flux in an unphysical way.  Second, it spuriously enhances the thermal emission.  We recommend that the hemispheric closure be used instead when computing synthetic spectra.

\item \textbf{The photosphere does not always reside at an optical depth of 2/3: } The ``2/3 rule" comes from Milne's solution for self-luminous atmospheres (stars).  We demonstrate that in a highly irradiated atmosphere with internal heat, the location of the photosphere (as computed from its corresponding value of the optical depth) depends on the relative strength of shortwave versus longwave absorption and the properties of scattering.

\end{itemize}

To supplement our current study, we derive total, net, outgoing and incoming fluxes in the convective regime in Appendix \ref{append:convective}, thus generalizing the work of \cite{robinson12}, who computed them in the purely absorbing limit.

\subsection{Comparison to Previous Analytical Work}

Overall, the novel aspect of our study is the construction of a unified, self-consistent framework for studying two-stream radiative transfer, flux-limited diffusion and temperature-pressure profiles using the same set of governing equations and closures and enforcing general energy conservation.

\subsubsection{Two-Stream Radiative Transfer}

Several differences from past studies are worth mentioning.  \cite{mw80} and \cite{gy89} do not include the Planck function term in their derivations, cf. their equations (10)--(13) and (8.156), respectively.  \cite{mw80} list other closures such as ``modified Eddington", ``modified quadrature" and ``Delta function", but these are based on expressing the scattering phase function as a series expansion in Legendre polynomials \citep{chandra60}; these closures make different assumptions for the integrals associated with ${\cal P}$.  For the quadrature closure, \cite{toon89} and \cite{pierrehumbert} disagree on their expressions for $\gamma_{\rm B}$: the former states it as $\gamma_{\rm B} = 2 \pi \bar{\mu} \left( 1 - \omega_0 \right)$, while the latter writes $\gamma_{\rm B} = \sqrt{3} \pi \bar{\mu} \left( 1 - \omega_0 \right)$; we have chosen to list and implement the latter as it correctly produces $f_\infty=1$.

A major difference with our derivation is that we have omitted the contribution from a ``direct beam" \citep{chandra60}, which is usually included as a term involving the incident stellar flux, diluted across height by essentially a transmission function involving its own directionality: a characteristic value of $\mu$, which we may write as $\bar{\mu}_\star \equiv \cos\theta_\star$; the quantity $\theta_\star$ is often called the ``zenith angle".  
As already mentioned, given the fact that the two-stream treatment is one-dimensional and the incoming stellar flux may be modeled using the boundary condition at the top of the atmosphere, we consider this additional term to be superfluous \citep{mw80}.  Given the other assumptions and simplifications associated with the two-stream approximation,\footnote{\cite{pierrehumbert} describes the two-stream approximations as ``what physicists euphemistically like to call `uncontrolled approximations,' in that they are not actually exact in any useful limit."} we find this approach to be reasonable.

\subsubsection{Analytical Temperature-Pressure Profiles}

Analytical temperature-pressure profiles for highly-irradiated atmospheres were first explored by \cite{hubeny03} and \cite{hansen08}.  \cite{guillot10} generalized these studies into a formalism describing both latitude-specific and globally-averaged temperature-pressure profiles, albeit in the limit of pure absorption.  \cite{hhps12} considered isotropic, coherent scattering, used non-constant infrared opacities and included a toy model for a Gaussian cloud deck, albeit with a constant optical opacity.  \cite{robinson12} augmented temperature-pressure profiles in the purely absorbing limit with convective adiabats and employed the diffusivity factor (see \S\ref{subsect:diffusivity}).  (See also Appendix \ref{append:convective}.) \cite{pg14} generalized the ``picket fence model", previously described in \cite{mihalas}, to describe highly-irradiated atmospheres, by including four opacities to mimic the presence of spectral lines and continua, albeit in the purely absorbing limit.

In the present study, we obtain solutions for a non-constant (power-law) optical opacity and consider non-isotropic, coherent scattering both in the optical and infrared.  We also distinguish between net and total fluxes, such that a heating term previously derived by \cite{guillot10} and \cite{hhps12} naturally vanishes when radiative equilibrium is attained.

\subsection{Relegating the Burden of Isothermality to Numerical Resolution?}
\label{subsect:burden}

A real atmosphere is described by a continuous temperature-pressure profile, which our two-stream model is trying to approximate as a collection of discrete layers.  Within each layer, the simplest approach is to assign to it only a single temperature, i.e., isothermality.    

More realistically, we expect each model layer to possess an intra-layer temperature gradient, which means that the blackbody flux should vary across the layer.  In situations where the temperature-pressure profile is known (e.g., via in-situ measurements), such an approach is reasonable and robust.  In exoplanetary atmospheres, where the temperature-pressure profile is a priori unknown, attempting to model intra-layer temperature variations is computationally akin to assuming a sub-grid model.  In our models with non-isothermal layers, we are assuming that intra-layer variations may be linearly interpolated.  

As an initial approach, we expect that the accuracy of the isothermal assumption should be the burden of the numerical resolution of the calculation---if one desires a better answer, one simply needs to specify more layers within a model atmosphere.  An infinitesimally thin atmospheric layer may always be described as being isothermal.  In practice, the use of non-isothermal layers is computationally efficient, leads to rapid numerical convergence and may be more accurate than using an equivalent number of isothermal layers \citep{lo91}.  In exoplanetary atmospheres, the advantages and disadvantages of using isothermal versus non-isothermal layers remain to be fully elucidated in future numerical work.

Deep in the interior, as the vertical resolution far exceeds the photon mean free path, the heating is more conveniently described by flux-limited diffusion, which is a demonstrably non-isothermal phenomenon (see \S\ref{subsect:fld}).

\subsection{The Diffusivity Factor}
\label{subsect:diffusivity}

In the two-stream approximation, one may generally write the transmission function as
\begin{equation}
{\cal T} = \exp{\left( - {\cal D} \Delta \tau_0 \right)},
\end{equation}
where ${\cal D}$ is often termed a ``diffusivity factor".  Within the context of our formalism, the diffusivity factor is related to the first Eddington coefficient,
\begin{equation}
{\cal D} = \frac{1}{\epsilon_\pm}.
\end{equation}
In order to produce the correct blackbody flux for a purely absorbing, isothermal, opaque atmosphere, we have set $\epsilon_\pm = 1/2$ or ${\cal D}=2$.  Making other choices for the value of ${\cal D}$ alters this asymptotic value of the blackbody flux ($4\pi B/{\cal D}$).

For a purely absorbing atmosphere, it is possible to solve for ${\cal D}$ by solving the radiative transfer equation directly for the intensity (see Appendix \ref{append:direct}), instead of using the method of moments.  The transmission function then takes on a generalized form, cf. equation (\ref{eq:transmission_general}), which requires an integration over $\mu$.  The diffusivity factor is then obtained by solving the equation,
\begin{equation}
\exp{\left(- {\cal D} \Delta \tau_0\right)} = 2 \int^1_0 \mu \exp{\left(-\frac{\Delta \tau_0}{\mu} \right)} ~d\mu.
\label{eq:solve_diffusivity}
\end{equation}
It is apparent that the value of ${\cal D}$ depends on $\Delta \tau_0$, which is the difference in the (non-slanted) optical depth between two atmospheric layers.  Since the radiative transfer equation cannot be solved directly when scattering is present (see Appendix \ref{append:direct}), one cannot write down a generalized form of equation (\ref{eq:solve_diffusivity}) and solve for ${\cal D}$ when $\omega_0 \ne 0$.

Numerically, an optimal value for the diffusivity factor may be inferred by performing a calculation based on the direct solution of the radiative transfer equation and comparing it to a second calculation based on the two-stream solutions with a chosen value of ${\cal D}$.  For example, \cite{amundsen14} report that ${\cal D}=1.66$ accurately approximates the direct solution, based on comparing several calculations of hot exoplanetary atmospheres in the purely absorbing limit.  Coincidentally, ${\cal D}=1.649$--1.66 is motivated by an Earth-centric calculation (mostly of water under Earth-like conditions) of the terrestrial atmosphere \citep{armstrong68,armstrong69}.

\acknowledgments
KH, JM and JL thank the Center for Space and Habitability (CSH) and the Space Research and Planetary Sciences Division (WP) of the University of Bern for financial, secretarial and logistical support.  KH and JL acknowledge partial financial support from the Swiss-based MERAC Foundation via a grant awarded to KH for the Exoclimes Simulation Platform.  KH acknowledges financial support from the Swiss National Science Foundation.  We are grateful to Ray Pierrehumbert for a useful conversation on the symmetry properties of the scattering phase function.  We thank the anonymous referee for fair, constructive and detailed reports that improved the quality and clarity of the manuscript.

\appendix

\section{Total, Net, Outgoing and Incoming Fluxes in the Convective Regime}
\label{append:convective}

We generalize the work of \cite{robinson12}, who computed fluxes in the convective regime, by including non-isotropic, coherent scattering.  The mathematical machinery for deriving these fluxes has already been laid out in \S\ref{subsect:general_solution_iso}, so we will simply state our results.  As reasoned by \cite{robinson12}, the convective part of an atmosphere has a temperature-pressure profile given by $T \propto P^{(\gamma-1)/\gamma}$, where $\gamma$ is the adiabatic gas index.  The index $(\gamma-1)/\gamma$ may be diluted by a factor of order unity to mimic moist convection.  Consider the opacity to be $\propto m^{n_0} \propto P^{n_0}$, where $n_0$ is a dimensionless index, which implies that $\tau \propto P^{n_0+1}$.  Here, we are using $\tau$ to represent the slant optical depth integrated over all wavelengths, although we expect most of its contributions to come from the longwave.  It follows that the temperature-pressure profile is given by
\begin{equation}
T = T_{\rm BOA} \left( \frac{\tau}{\tau_{\rm BOA}} \right)^{\left( \gamma - 1 \right)/\gamma\left(n_0+1\right)},
\label{eq:tconvect}
\end{equation}
where $T_{\rm BOA}$ and $\tau_{\rm BOA}$ are the temperature and slant optical depth, respectively, at the bottom of the atmosphere, along this convective adiabat.  Note that $T_{\rm BOA} \ne T_{\rm s}$ in general (e.g., \citealt{pierrehumbert}), since the surface of an exoplanet may be hotter than the atmosphere directly above it.

By setting
\begin{equation}
n_{\rm c} \equiv \frac{4 \left( \gamma - 1 \right)}{\gamma\left( n_0 + 1 \right)}, ~\alpha_{\rm c} \equiv \frac{2 \gamma_{\rm B} \left( \gamma_{\rm a} + \gamma_{\rm s} \right) \sigma_{\rm SB} T^4_{\rm BOA}}{\pi \tau_{\rm BOA}^{n_{\rm c}}},
\end{equation}
the governing equation for the total flux, integrated over all wavelength, becomes
\begin{equation}
\frac{\partial^2 {\cal F}_+}{\partial \tau^2} - \alpha^2 {\cal F}_+ + \alpha_{\rm c} \tau^{n_{\rm c}} = 0,
\end{equation}
where $\alpha$ has previously been defined in equation (\ref{eq:alpha_2}).

The total and net fluxes are
\begin{equation}
\begin{split}
{\cal F}_+ =& {\cal A}_1 \exp{\left(\alpha \tau \right)} + {\cal A}_2 \exp{\left(-\alpha \tau \right)} - \frac{\alpha_{\rm c} \tau^{n_{\rm c}+2}}{\left(n_{\rm c}+1\right)\left(n_{\rm c}+2\right)}, \\
{\cal F}_- =& \left( \frac{\gamma_{\rm a} - \gamma_{\rm s}}{\gamma_{\rm a} + \gamma_{\rm s}} \right)^{1/2} \left[ {\cal A}_1 \exp{\left(\alpha \tau \right)} - {\cal A}_2 \exp{\left(-\alpha \tau \right)} \right] - \frac{\alpha_{\rm c} \tau^{n_{\rm c}+1}}{\left(n_{\rm c}+1\right)\left( \gamma_{\rm a} + \gamma_{\rm s} \right)}.
\end{split}
\end{equation}
From these expressions, we may derive ${\cal F}_\uparrow$ and ${\cal F}_\downarrow$, albeit with the coefficients ${\cal A}_1$ and ${\cal A}_2$ still present.  To eliminate them requires enforcing the boundary conditions ${\cal F}_{\uparrow_2}$ and ${\cal F}_{\downarrow_1}$.  For a pair of atmosphere layers (where $\tau_1 < \tau_2$), the outgoing and incoming fluxes are
\begin{equation}
\begin{split}
{\cal F}_{\uparrow_1} =& \frac{1}{\left( \zeta_- {\cal T} \right)^2 - \zeta_+^2} \left\{ \left( \zeta_-^2 - \zeta_+^2 \right) {\cal T} {\cal F}_{\uparrow_2} - \zeta_- \zeta_+ \left( 1 - {\cal T}^2 \right) {\cal F}_{\downarrow_1} \right. \\
&+ \left. \frac{\alpha_{\rm c}}{\left(n_{\rm c}+1\right)\left( \gamma_{\rm a}+\gamma_{\rm s} \right)} \left[ {\cal T} \tau_2^{n_{\rm c}+1} \left( \zeta_-^2 \left[ 2 {\cal T}^2 - 1 \right] - \zeta^2_+ \right) + \tau_1^{n_{\rm c}+1} \left( \zeta_+ - \zeta_- \right) \left( \zeta_+ + \zeta_- {\cal T}^2 \right) \right] \right. \\
&+ \left. \frac{\alpha_{\rm c}}{\left( n_{\rm c}+1 \right)\left( n_{\rm c}+2 \right)} \left[ {\cal T} \tau_2^{n_{\rm c}+2} \left( \zeta_-^2 \left[ 2 {\cal T}^2 - 1 \right] - \zeta^2_+ \right) + \tau_1^{n_{\rm c}+2} \left( \zeta_- + \zeta_+ \right) \left( \zeta_+ - \zeta_- {\cal T}^2 \right) \right] \right\},  \\
{\cal F}_{\downarrow_2} =& \frac{1}{\left( \zeta_- {\cal T} \right)^2 - \zeta_+^2} \left\{ \left( \zeta_-^2 - \zeta_+^2 \right) {\cal T} {\cal F}_{\downarrow_1} - \zeta_- \zeta_+ \left( 1 - {\cal T}^2 \right) {\cal F}_{\uparrow_2} \right. \\
&+ \left. \frac{\alpha_{\rm c}}{\left(n_{\rm c}+1\right)\left( \gamma_{\rm a}+\gamma_{\rm s} \right)} \left[ \tau_2^{n_{\rm c}+1} \left( \zeta_- - \zeta_+ \right) \left( \zeta_+ + \zeta_- {\cal T}^2 \right) + {\cal T} \tau_1^{n_{\rm c}+1} \left( \zeta_+^2 \left[ 2 {\cal T}^{-2} - 1 \right] - \zeta_-^2  \right) \right] \right. \\
&+ \left. \frac{\alpha_{\rm c}}{\left(n_{\rm c}+1\right)\left( n_{\rm c}+2 \right)} \left[ \tau_2^{n_{\rm c}+2} \left( \zeta_- + \zeta_+ \right)\left( \zeta_+ - \zeta_- {\cal T}^2 \right) + {\cal T} \tau_1^{n_{\rm c}+2} \left( \zeta_-^2 - \zeta_+^2 \left[ 2 {\cal T}^{-2} - 1 \right] \right) \right] \right\}.  \\
\end{split}
\label{eq:fluxes_convect}
\end{equation}

Consider an atmosphere where the convective region sits below some depth, at $\tau \ge \tau_{\rm c}$, where the transition (slant) optical depth ($\tau_{\rm c}$) may be computed by equating $T$ in equation (\ref{eq:tconvect}) to $\bar{T}$ in equation (\ref{eq:tp_global}).  To use the equations in (\ref{eq:fluxes_convect}) in the same way as in \cite{robinson12}, one has to set ${\cal F}_{\uparrow_2} = \sigma_{\rm SB} T^4_{\rm BOA}$, $\tau_2 = \tau_{\rm BOA}$ and $\tau_1 = \tau$.  The other boundary condition is ${\cal F}_{\downarrow_1} = \sigma_{\rm SB} T^4_{\rm c}$, where $T_{\rm c} \equiv T(\tau_{\rm c})$.  Note that since we can never have ${\cal T} = 0$ when the equations in (\ref{eq:fluxes_convect}) are used in this way, ${\cal F}_{\downarrow_2}$ will not diverge due to the ${\cal T}^{-2}$ terms.  

Unlike in the purely absorbing case, as found by \cite{robinson12}, the outgoing and incoming fluxes depend on both boundary conditions in the presence of scattering.  Furthermore, we have circumvented the need to use incomplete gamma functions, as was the approach in \cite{robinson12}, by solving a second-order differential equation for ${\cal F}_+$, instead of a first-order one (see \S\ref{subsect:equivalence}).  

It is worth noting that the equations in (\ref{eq:fluxes_convect}) lack the symmetry of those in (\ref{eq:twostream_iso_general_solution_non_isothermal}) (between $F_{\uparrow_1}$ and $F_{\downarrow_2}$), because we have enforced a temperature-pressure profile that is asymmetric across pressure or height.

While we have discussed the use of the equations in (\ref{eq:fluxes_convect}) for the convective part of the atmosphere just above the surface of an exoplanet, they may also be used to describe detached convective regions.

\section{Direct Analytical Solution of the Radiative Transfer Equation and Why It Only Works for Pure Absorption}
\label{append:direct}

In the limit of pure absorption, the radiative transfer equation may be solved directly for the intensity, circumventing the need for the method of moments.  However, such an approach breaks down when scattering is present.  To demonstrate this, we assume isotropic, coherent scattering, as described by equation (\ref{eq:rt_isotropic}), and obtain
\begin{equation}
I_2 \exp{\left( -\frac{\tau_{0_2}}{\mu} \right)} - I_1 \exp{\left( -\frac{\tau_{0_1}}{\mu} \right)} = - \frac{1}{\mu} \int^{\tau_{0_2}}_{\tau_{0_1}} \left[ \frac{\omega_0 J}{4\pi} + \left( 1 - \omega_0 \right) B \right] \exp{\left( -\frac{\tau_0}{\mu} \right)} ~d\tau_0,
\label{eq:direct}
\end{equation}
where $I_2$ and $I_1$ are the intensities evaluated at $\tau_0 = \tau_{0_2}$ and $\tau_0 = \tau_{0_1}$, respectively.  When $\omega_0 \ne 0$, the integral cannot be evaluated since the functional form of $J$ is a priori unknown.  It cannot be assumed that $J$ obeys isothermality (i.e., is independent of $\tau_0$), because it is related to the outgoing and incoming fluxes via an Eddington coefficient and the fluxes generally depend on $\tau_0$.

However, when $\omega_0=0$, we may evaluate equation (\ref{eq:direct}) for isothermal atmospheric layers,
\begin{equation}
I_1 = I_2 {\cal T}_0 + B \left( 1 - {\cal T}_0 \right),
\label{eq:direct2}
\end{equation}
where we have defined
\begin{equation}
{\cal T}_0 \equiv \exp{\left( - \frac{ \Delta \tau_0 }{\mu} \right)} 
\end{equation}
and $\Delta \tau_0 \equiv \tau_{0_2} - \tau_{0_1} > 0$.  By assuming $I_1$ and $I_2$ to be constant with respect to $\mu$ and $\phi$, one multiplies equation (\ref{eq:direct2}) by $\mu$, integrates over $d\Omega = d\mu d\phi$ in each hemisphere and obtains
\begin{equation}
\begin{split}
F_{\uparrow_1} &= F_{\uparrow_2} {\cal T} + \pi B \left( 1 - {\cal T} \right), \\
F_{\downarrow_2} &= F_{\downarrow_1} {\cal T} + \pi B \left( 1 - {\cal T} \right), \\
\end{split}
\end{equation}
if we identify $F_{\uparrow\downarrow_1} = \pi I_1$ and $F_{\uparrow\downarrow_2} = \pi I_2$.  The transmission function now takes on a more general form,
\begin{equation}
{\cal T} \equiv 2 \int^1_0 \mu \exp{\left( -\frac{\Delta \tau_0}{\mu} \right)} ~d\mu = \left( 1 - \Delta \tau_0 \right) \exp{\left(-\Delta \tau_0 \right)} + \left( \Delta \tau_0 \right)^2 {\cal E}_1,
\label{eq:transmission_general}
\end{equation}
with ${\cal E}_1 = {\cal E}_1( \Delta \tau_0 )$ being the exponential integral of the first order.  It is important to note that this generalized form of ${\cal T}$ is only valid in the limit of pure absorption.

If we express the Planck function as given by equation (\ref{eq:b_expand}), then the direct solutions become
\begin{equation}
\begin{split}
F_{\uparrow_1} &= F_{\uparrow_2} {\cal T} + \pi B_2 \left( 1 - {\cal T} \right) + \pi B^\prime \left\{ \frac{2}{3} \left[ 1 - \exp{\left( -\Delta \tau_0 \right)} \right] - \Delta \tau_0 \left( 1 - \frac{{\cal T}}{3} \right) \right\}, \\
F_{\downarrow_2} &= F_{\downarrow_1} {\cal T} + \pi B_1 \left( 1 - {\cal T} \right) + \pi B^\prime \left\{ -\frac{2}{3} \left[ 1 - \exp{\left( -\Delta \tau_0 \right)} \right] + \Delta \tau_0 \left( 1 - \frac{{\cal T}}{3} \right) \right\}. \\
\end{split}
\end{equation}


\label{lastpage}


\begin{thebibliography}{99}

\bibitem[Abramowitz \& Stegun(1970)]{abram} Abramowitz, M., \& Stegun, I.A. \ 1970, Handbook of Mathematical Functions, 9th printing (New York: Dover Publications)

\bibitem[Amundsen et al.(2014)]{amundsen14} Amundsen, D.S., Baraffe, I., Tremblin, P., Manners, J., Hayek, W., Mayne, N.J., \& Acreman, D.M. \ 2014, A\&A, 564, A59

\bibitem[Angerhausen et al.(2014)]{angerhausen14} Angerhausen, D., DeLarme, E., \& Morse, J.A. \ 2014, arXiv:1404.4348v1

\bibitem[Arfken \& Weber(1995)]{aw95} Arfken, G.B., \& Weber, H.J. \ 1995, Mathematical Methods for Physicists, 4th edition (San Diego: Academic Press)

\bibitem[Armstrong(1968)]{armstrong68} Armstrong, B.H. \ 1968, Journal of Quantitative Spectroscopy and Radiative Transfer, 8, 1577

\bibitem[Armstrong(1969)]{armstrong69} Armstrong, B.H. \ 1969, Journal of the Atmospheric Sciences, 26, 741

\bibitem[Benneke \& Seager(2012)]{benneke12} Benneke, B., \& Seager, S. \ 2012, ApJ, 753, 100

\bibitem[Burrows et al.(2008)]{burrows08} Burrows, A., Budaj, J., \& Hubeny, I. \ 2008, ApJ, 678, 1436

\bibitem[Chandrasekhar(1960)]{chandra60} Chandrasekhar, S. \ 1960, Radiative Transfer (New York: Dover Publications)

\bibitem[Draine \& Lee(1984)]{draine84} Draine, B.T., \& Lee, H.M. \ 1984, ApJ, 285, 89

\bibitem[Evans et al.(2013)]{evans13} Evans, T.M. \ 2013, ApJ, 772, L16

\bibitem[Fortney et al.(2010)]{fortney10} Fortney, J.J., Shabram, M., Showman, A.P., Lian, Y., Freedman, R.S., Marley, M.S., \& Lewis, N.K. \ 2010, ApJ, 709, 1396

\bibitem[Frierson, Held \& Zurita-Gotor(2006)]{fhz06} Frierson, D.M.W., Held, I.M., \& Zurita-Gotor, P. \ 2006, Journal of the Atmospheric Sciences, 63, 2548

\bibitem[Goody \& Yung(1989)]{gy89} Goody, R.M., \& Yung, Y.L. \ 1989, Atmospheric Radiation: Theoretical Basis, 2nd edition (New York: Oxford University Press)

\bibitem[Guillot(2010)]{guillot10} Guillot, T. \ 2010, A\&A, 520, A27

\bibitem[Hansen(2008)]{hansen08} Hansen, B.M.S. \ 2008, ApJS, 179, 484

\bibitem[Heng, Menou \& Phillipps(2011)]{hmp11} Heng, K., Menou, K., \& Phillipps, P.J. \ 2011, MNRAS, 413, 2380

\bibitem[Heng et al.(2012)]{hhps12} Heng, K., Hayek, W., Pont, F., \& Sing, D.K. \ 2012, MNRAS, 420, 20

\bibitem[Heng \& Kopparla(2012)]{hk12} Heng, K., \& Kopparla, P. \ 2012, ApJ, 754, 60

\bibitem[Heng \& Demory(2013)]{hd13} Heng, K., \& Demory, B.-O. \ 2013, ApJ, 777, 100

\bibitem[Heng \& Workman(2014)]{hw14} Heng, K., \& Workman, J. \ 2014, ApJS, 213, 27

\bibitem[Hubeny et al.(2003)]{hubeny03} Hubeny, I., Burrows, A., \& Sudarsky, D. \ 2003, ApJ, 594, 1011

\bibitem[Ingersoll(1969)]{ingersoll69} Ingersoll, A.P. \ 1969, Journal of the Atmospheric Sciences, 26, 1191

\bibitem[Komabayashi(1967)]{koma67} Komabayashi, M. \ 1967, Journal of the Meteorological Society of Japan, 45, 137

\bibitem[Lacis \& Oinas(1991)]{lo91} Lacis, A.A., \& Oinas, V. \ 1991, Journal of Geophysical Research, 96, 9027

\bibitem[Laor \& Draine(1993)]{laor93} Laor, A., \& Draine, B.T. \ 1993, ApJ, 402, 441

\bibitem[Lee et al.(2012)]{lee12} Lee, J.-M., Fletcher, L.N., \& Irwin, P.G.J. \ 2012, MNRAS, 420, 170

\bibitem[Levermore \& Pomraning(1981)]{lp81} Levermore, C.D., \& Pomraning, G.C. \ 1981, ApJ, 248, 321

\bibitem[Line et al.(2013)]{line13} Line, M.R., et al. \ 2013, ApJ, 775, 137

\bibitem[Madhusudhan \& Burrows(2012)]{mb12} Madhusudhan, N., \& Burrows, A. \ 2012, ApJ, 747, 25

\bibitem[Manabe, Smagorinsky \& Strickler(1965)]{manabe65} Manabe, S., Smagorinsky, J., \& Strickler, R.F. \ 1965, Monthly Weather Review, 93, 769

\bibitem[Marley et al.(1999)]{marley99} Marley, M.S., Gelino, C., Stephens, D., Lunine, J.I., \& Freedman, R. \ 1999, ApJ, 513, 879

\bibitem[Meador \& Weaver(1980)]{mw80} Meador, W.E., \& Weaver, W.R. \ 1980, Journal of the Atmospheric Sciences, 630, 37

\bibitem[Mihalas(1970)]{mihalas} Mihalas, D. \ 1970, Stellar Atmospheres (San Francisco: Freeman)

\bibitem[Mihalas \& Weibel-Mihalas(1999)]{mm99} Mihalas, D., \& Weibel-Mihalas, B. \ 1999, Foundations of Radiation Hydrodynamics, 2nd Edition (New York: Dover Publications)

\bibitem[Narayan(1992)]{narayan92} Narayan, R. \ 1992, ApJ, 394, 261

\bibitem[Parmentier \& Guillot(2014)]{pg14} Parmentier, V., \& Guillot, T. \ 2014, A\&A, 562, A133

\bibitem[Pierrehumbert(2010)]{pierrehumbert} Pierrehumbert, R.T. \ 2010, Principles of Planetary Climate (New York: Cambridge University Press)

\bibitem[Rauscher \& Menou(2012)]{rm12} Rauscher, E., \& Menou, K. \ 2012, ApJ, 750, 96

\bibitem[Robinson \& Catling(2012)]{robinson12} Robinson, T.D., \& Catling, D.C. \ 2012, ApJ, 757, 104

\bibitem[Russell(1916)]{russell16} Russell, H.N. \ 1916, ApJ, 43, 173

\bibitem[Seager(2010)]{seager10} Seager, S. \ 2010, Exoplanet Atmospheres (New Jersey: Princeton University Press)

\bibitem[Showman et al.(2009)]{showman09} Showman, A.P., Fortney, J.J., Lian, Y., Marley, M.S., Freedman, R.S., Knutson, H.A., \& Charbonneau, D. \ 2009, ApJ, 699, 564

\bibitem[Toon et al.(1989)]{toon89} Toon, O.B., McKay, C.P., \& Ackerman, T.P. \ 1989, Journal of Geophysical Research, 94, 16287

\end{thebibliography}
\end{document}